\documentclass[superscriptaddress,twocolumn,10pt,pra,aps,showpacs,longbibliography]{revtex4-1}
\usepackage{float}
\floatstyle{boxed}
\usepackage{wrapfig}
\usepackage{array}
\usepackage{graphicx}
\usepackage{amsbsy}
\usepackage[utf8x]{inputenc}
\usepackage{epstopdf}
\usepackage{amsmath,amssymb,amsfonts}
\usepackage{dsfont}
\usepackage{cprotect}

\usepackage[dvipsnames]{xcolor}

\def \am{\hat a}
\def \ap{\hat a^{\dagger}}

\newcommand{\de}{{\rm d}}

\newcommand{\bra}[1]{\langle #1|}
\newcommand{\ket}[1]{|#1\rangle}

\newcommand{\expec}[1]{\left\langle #1 \right\rangle}

\newcommand{\im}{\text{Im}}
\newcommand{\re}{\text{Re}}

\renewcommand{\eqref}[1]{\mbox{Eq.~(\ref{#1})}}
\renewcommand{\Re}[1]{{\rm Re}\left[#1 \right]}

\newcommand{\be}{\begin{equation}}
\newcommand{\ee}{\end{equation}}
\newcommand{\bea}{\begin{eqnarray}}
\newcommand{\eea}{\end{eqnarray}}

\usepackage{multirow}

\usepackage{url}
\usepackage[colorlinks]{hyperref}
\hypersetup{
    plainpages=true,
    breaklinks=true,
    hypertexnames=false,
    pageanchor=true,
    colorlinks=true,
    linkcolor={blue},
    citecolor={red},
    urlcolor={blue},
    anchorcolor={black}
}

\newcommand{\LL}{\mathcal{L}}
\newcommand{\DD}{\mathcal{D}}
\newcommand{\rhot}{\hat{\rho}(t)}
\newcommand{\sss}{\hat{\rho}_{\rm ss}}
\newcommand{\eig}[1]{\hat{\rho}_{#1}}

\makeatletter
\newcommand*\bigcdot{\mathpalette\bigcdot@{.5}}
\newcommand*\bigcdot@[2]{\mathbin{\vcenter{\hbox{\scalebox{#2}{$\m@th#1\bullet$}}}}}
\makeatother

\usepackage{enumerate}
\usepackage[normalem]{ulem}

\begin{document}

\author{Fabrizio Minganti }
\email{fabrizio.minganti@gmail.com} 
\affiliation{Theoretical
Quantum Physics Laboratory, RIKEN Cluster for Pioneering Research,
Wako-shi, Saitama 351-0198, Japan}
\affiliation{Institute of Physics, Ecole Polytechnique F\'ed\'erale de Lausanne (EPFL), CH-1015 Lausanne, Switzerland}
\author{Ievgen I. Arkhipov}
\email{ievgen.arkhipov@upol.cz} \affiliation{Joint Laboratory of
Optics of Palack\'y University and Institute of Physics of CAS,
Faculty of Science, Palack\'y University, 17. listopadu 12, 771 46
Olomouc, Czech Republic}
\author{Adam Miranowicz}
\email{miran@amu.edu.pl} \affiliation{Theoretical Quantum Physics
Laboratory, RIKEN Cluster for Pioneering Research, Wako-shi,
Saitama 351-0198, Japan} \affiliation{Institute of Spintronics and Quantum Information,
Faculty of Physics, Adam Mickiewicz University, 61-614 Pozna\'n, Poland}
\author{Franco Nori}
\email{fnori@riken.jp}
 \affiliation{Theoretical Quantum Physics
Laboratory, RIKEN Cluster for Pioneering Research, Wako-shi,
Saitama 351-0198, Japan} \affiliation{Physics Department, The
University of Michigan, Ann Arbor, Michigan 48109-1040, USA}

\title{Liouvillian spectral collapse in the Scully-Lamb laser model}

\begin{abstract}
Phase transitions of thermal systems and the laser threshold were first connected more than forty years ago.
Despite the nonequilibrium nature of the laser, the Landau theory of thermal phase transitions, applied directly to the Scully-Lamb laser model (SLLM), indicates that the laser threshold is a second-order phase transition, associated with a $U(1)$ spontaneous symmetry breaking (SSB).
To capture the genuine 
{\it nonequilibrium} phase transition of the SLLM (i.e., a single-mode laser without a saturable absorber), here we employ a quantum theory of dissipative phase transitions. 
Our results confirm that the $U(1)$ SSB can occur at the lasing threshold but, in contrast to the Landau theory and semiclassical approximation, they signal that the SLLM ``fundamental'' transition is a different phenomenon, which we call \textit{Liouvillian spectral collapse}; that is, the emergence of diabolic points of infinite degeneracy.
By considering a \textit{generalized SLLM} with additional dephasing, we witness a second-order phase transition, with a Liouvillian spectral collapse, but in the absence of symmetry breaking. 
Most surprisingly, the phase transition corresponds to the emergence of dynamical multistability even without SSB. 
Normally, bistability is suppressed by quantum fluctuations, while in this case, the very presence of quantum fluctuations enables bistability.
This rather anomalous bistability, characterizing the truly dissipative and quantum origin of lasing, can be an experimental signature of our predictions, and we show that it is associated with an emergent dynamical hysteresis.
\end{abstract}

\date{\today}

\maketitle

\tableofcontents


\section{Introduction}

The study and experimental realization of the lasing transition is a milestone in quantum optics. The original idea of population inversion in terms of Einstein's emission and absorption
coefficients \cite{Einstein16}, paved the way to the
creation of devices and lasers exploiting the stimulated emission of optical radiation \cite{MAIMAN1960}. The Scully-Lamb laser model (SSLM) is a remarkably predictive quantum description of a laser oscillator~\cite{Scully1967}. 
Recently, the SLLM has attracted much interest in different fields, ranging from open quantum systems phenomena
(e.g., exceptional points~\cite{Ozdemir2019,Arkhipov2019} and
quantum thermodynamics~\cite{BinderBook,Scully2019}) to more
exotic ones (the characterization of entropy in Bose-Einstein
condensates and even in black holes~\cite{Scully2019}).

As pointed out in, e.g., Ref.~\cite{Landau_BOOK_Statistical}, the laser population inversion can never occur in thermal systems, since it
would imply the presence of negative temperatures \cite{Braun13}.
As such, the laser is an out-of-equilibrium system \cite{Sun2008}.
Despite this fact, analogies between thermal phase transitions and lasing nonequilibrium transitions were drawn more
than forty years ago~\cite{Graham1970,DeGiorgio1970} using the Landau theory of phase
transitions~\cite{Landau1937,Landau_BOOK_Statistical,Landau_collection1965,Landau1936}. Accordingly, the SLLM threshold is a second-order phase transition associated with the spontaneous
breaking of the $U(1)$ symmetry~\cite{DeGiorgio1970}, and these results are supported by a \textit{semiclassical} analysis.

Notwithstanding the success of the Landau and semiclassical models in providing a qualitative description of lasing, still they lack a rigorous description of its transition. Indeed, the semiclassical description includes dissipative processes but neglects quantum fluctuations. The Landau theory takes into account thermal-like fluctuations but, by assuming thermodynamic equilibrium, neglects the out-of-equilibrium fluctuations \cite{BreuerBookOpen,LidarLectureNotes}.

As such, it is natural to ask if there is some unveiled physics behind the lasing process resulting from a genuine theory of quantum dissipative phase transitions, i.e., based on the quantum Liouvillian framework~\cite{MingantPRA18_Spectral}. 
Here, 
using the Lindblad master equation and its associated Liouvillian superoperator~\cite{MingantPRA18_Spectral}, we analyze the SLLM 
explicitly taking into consideration the $U(1)$ symmetry of the model. We divide the Liouvillian into its symmetry sectors, i.e., we separate the dynamics into families of states which evolve independently one from another due to the presence of the $U(1)$ symmetry \cite{AlbertPRA14,BucaNPJ2012,PalacinoPRR21}. 
\textit{We demonstrate that the SLLM second-order phase transition is characterized by a Liouvillian spectral collapse}. That is, a high degeneracy of the Liouvillian spectrum emerges as diabolic points of infinite order, one for each symmetry sectors, triggering dynamical hysteresis and other critical properties which cannot be explained by the $U(1)$ SSB alone. 
We borrow the term spectral collapse from the two-photon Rabi model---describing a parametric exchange of excitations between a bosonic field and an ensemble of spins---where high-order degeneracies of the Hamiltonian emerge at the critical coupling (see the discussion in, e.g., Refs.~\cite{FelicettiPRA15,GarbePRA17}). Although similar, this Liouvillian spectral collapse is not a trivial extension of its Hamiltonian counterpart, but rather a \textit{novel type of criticality}.    
Using the theory recently developed in Ref.~\cite{Minganti2021continuous}, we give a more insightful description of
 the nature of the spectral collapse by preventing spontaneous symmetry breaking (SSB) from the SSLM. 
To do that, we consider a \textit{generalized} SLLM, i.e., the standard SLLM model with constant additional dephasing. 
While the $U(1)$ symmetry is maintained, long-lived phase coherences are destroyed, thus preventing the emergence of a symmetry-broken phase and highlighting some properties which distinctively characterize the spectral collapse.

It is worth mentioning that, according to (semi)classical
equations of motion, a bifurcation phenomenon in equilibrium
systems can also be accompanied by a second- (third-) order phase
transition with (without) symmetry
breaking~\cite{Hachisu1983,CONSTANTINESCU1979,Ekiz2004}.
Moreover, in thermal systems, a second-order phase transition
without symmetry breaking may also occur, as can be described in a
topological-transition framework~\cite{Kubo2013}. However, the
SSLM with the removed SSB \textit{always has a unique steady
state}, as shown in Ref.~\cite{Minganti2021continuous}. This
property prohibits the occurrence of semiclassical bifurcations
and multistabilities, which, in turn, demonstrates that its nature
is profoundly different from that of the known effects.

Notably, some pioneering works grasped hints of the emergence of a critical timescale in the laser not associated with SSB. Indeed, in Ref.~\cite{Risken1967} the presence of a slow timescale was argued using a potential-well approximation to compute the tunneling time in a Fokker-Planck formalism. 
Moreover, in Ref.~\cite{WangPRA73}, 
it is explicitly shown that there is a closure of the spectral gap (to be associated with the emergence of a critical timescale in the photon-number evolution) in the limit of a vanishing saturation rate. These results corroborate the validity of the Liouvillian analysis of the SLLM transition.

An analogous effect of a \textit{mixed-order phase transition}
was described in Ref.~\cite{HuberPRA20}, where it was shown that a
spin model, also characterized by a $U(1)$ symmetry, can undergo a
dissipative phase transition without SSB. Furthermore, a
Liouvillian spectral collapse, similar to the one studied here,
has also been observed in more exotic lasing models, such as the
squeezed laser in Ref.~\cite{Munozarxiv20}, or in a $U(1)$
symmetric superradiant model~\cite{PalacinoPRR21}. In this regard,
the model under consideration exhibits behavior similar to that in
these other $U(1)$-symmetric systems. With the help of our
analytical description via the $P$-function, we can precisely
estimate the nature of quantum fluctuations, which induce this
peculiar character of the transition. Note also that the SSB in
our model can be arbitrarily removed according to the theory
developed in Ref.~\cite{Minganti2021continuous}. This approach
enables us to describe more deeply the interplay between the
$U(1)$ symmetry and quantum fluctuations in triggering critical
timescales.

In this regard, our analysis shows a novel behavior of dissipative phase transitions \cite{KesslerPRA12,JingPRL14,MarinoPRL2016,RotaPRL19,SorientePRR21,ArkadevNatComm21,Rossiniarxiv21}.
Indeed, while hysteresis \cite{CasteelsPRA16,RodriguezPRL17,LandaPRL20} and slowing-down properties \cite{CasteelsPRA17-2,FinkPRX17,FinkNatPhys18} have been characterized for first-order dissipative transitions, here we demonstrate a dynamical hysteresis for a second-order phase transition with or without SSB. 
Furthermore, SSB in open systems has been discussed in, e.g., Refs. \cite{LeePRL13,BartoloPRA16,SavonaPRA17}, and  the role of $U(1)$ non-Hermitian processes, triggering the transition, have been characterized \cite{BiellaPRA17,TakemuraJOSAB21,Munozarxiv20}. The relation between criticality, symmetries, and exotic effects have also been discussed for a wide range of models \cite{BartoloEPJST17,RotaNJP18,HannukainenPRA18,MunozPRA19,TindallNJP20,MingantiarXiv20,HuberPRA20}.

The article is organized as follows. 
In Sec.~\ref{Sec:Model}, we recall the SLLM, and we introduce its generalization in the Liouvillian description. We recall the results of the semiclassical theory, and we derive the quantum solution in the weak-gain-saturation regime using the Glauber-Sudarshan $P$-representation. 
We discuss the connection between dephasing and the convergence rate to the steady state, and connect the $P$-function to a thermodynamic potential within the Landau theory.
In Sec~\ref{Sec:Intro_Spectral_collapse}, we demonstrate the one of the main results of this article, namely the presence of a Liouvillian spectral collapse in the SLLM. To do that, we consider the standard SLLM (i.e., without additional dephasing). Exploiting the Liouvillian symmetries, we confirm the presence of SSB within the Liouvillian theory.
Nevertheless, we show the presence of another critical phenomenon, not captured by the Landau theory, i.e., a Liouvillian spectral collapse at the critical point. 
In Sec.~\ref{Sec:Decoherence} we investigate the generalized SLLM, i.e., the standard SLLM in the presence of additional dephasing. 
As proved in Ref.~\cite{Minganti2021continuous}, this leads to a second-order phase transition in the absence of SSB. 
Notably, both the semiclassical theory and the Landau approach fail to describe this model. 
Hence, we validate the necessity to consider a Liouvillian theory to properly characterize the lasing transition.
We discuss the emergence of multistability at the critical point, witnessed by a dynamical hysteresis.
In Sec.~\ref{Sec:Characterization}, we detail the novelty of the spectral collapse, as well as similarities and differences with respect to other phase transitions. We characterize the nature of the fluctuations at the critical point, showing that the critical slowing down of the spectral collapse can be interpreted as the time required for a random process with constant diffusion to cover a larger and larger system size.
In Sec.~\ref{Sec:Conclusions}, we discuss the results obtained and
provide conclusions and perspectives. 
Appendix
\ref{Sec:Liouvillian_theory_full} details the procedure to obtain
the Liouvillian form of the SLLM equation of motion, as well as
its limits of validity. Moreover, it provides additional details on
the properties of the SLLM Liouvillian. 
Finally, in Appendix~\ref{Sec:Quantum_Fluctuations}, we provide additional
details concerning the lack of SSB in the SLLM using quantum trajectories.

\section{The Scully-Lamb model and its generalization}
\label{Sec:Model}

The time evolution of a quantum system under the Born and Markov approximation \cite{BreuerBookOpen} is captured by the Lindblad master equation
($\hbar=1$):
\begin{equation}\label{Eq:Lindblad1}
\frac{\de}{\de t}\rhot={\cal L}\hat\rho(t)=-i\left[\hat
H,\rhot\right] +\sum\limits_{j} \DD[\hat{L}_j]\rhot,
\end{equation}
where $\rhot$ is the system reduced density matrix at time $t$, $\hat{H}$ is the Hamiltonian describing the coherent part of the system evolution, $\LL$ is the Liouvillian
superoperator~\cite{Carmichael_BOOK_2}, and $\DD[L_j]$ are the
so-called Lindblad dissipators, whose action is
\begin{equation}\label{Eq:Dissipator_superoperator}
    \DD[\hat{L}_j]\rhot=\hat L_i\rhot\hat L_j^{\dagger} - \frac{\hat L_j^{\dagger}\hat L_j\rhot+\rhot\hat L_j^{\dagger}\hat L_j}{2}.
\end{equation}
The operators $\hat{L}_j$ are the jump operators, which
describe how the environment acts on the system inducing loss and
gain of particles, energy, and information. A central role in the
following discussion is played by the steady state $\sss$,
i.e., the state which does not evolve any more under the action of
the Liouvillian: 
\begin{equation}
    \partial_t \sss=\LL \sss = 0.
\end{equation}
We indicate the expectation values of operators in the steady state as $\expec{\hat{o}}_{\rm ss} =\operatorname{Tr}[\sss \hat{o}]$.

\subsection{The model}

We consider here a quantum description of the Scully-Lamb laser model, describing a photonic mode pumped by the injection of inverted
atoms into the laser cavity~\cite{Walls_BOOK_quantum}. By tracing
out the atomic degrees of freedom~\cite{YamamotoBook}, the system
Hamiltonian becomes that of an optical cavity (a harmonic oscillator)
\begin{equation}\label{Eq:Hamiltonian}
    \hat{H}=\omega \hat{a}^\dagger \hat{a},
\end{equation}
where $\hat{a}$ ($\hat{a}^\dagger$) is the bosonic annihilation
(creation) operator. The evolution of the photonic field is
captured by the three jump operators (see the discussion in
Appendix~\ref{Sec:Liouvillian_theory_full} and
Refs.~\cite{WangPRA73,Gea1998,HenkelJPB07,ScullyLambBook,Arkhipov2019,ArkhipovPRA20,OrszagBook,YamamotoBook}):
\begin{subequations}\label{Eq:L123}  
\begin{equation}
\hat L_1 = \ap\left(\sqrt{A}-\frac{B}{2\sqrt{A}}\am\ap\right), 
\end{equation}
\begin{equation}
\hat L_2 =\sqrt{\beta}\am\ap= \sqrt{\frac{3B+4\eta}{4}}\am\ap,
\end{equation}
\begin{equation}
\hat L_3 =\sqrt{\Gamma}\am,
\end{equation}
\end{subequations}
where $\hat{L}_1$ describes the laser gain, $\hat{L}_2$ captures
the field dephasing, and $\hat L_3$ represents the particle loss. The
jump operators are characterized by rates: $A$
for the \emph{unsaturated gain} of the active medium~\cite{Sun2008} (we are considering here a laser without an absorber), $B$ for the gain saturation, $\Gamma$ for the dissipation rate, which corresponds to the inverse of the photon lifetime. 
The rate $\beta$ represents the overall dephasing rate of the system; accordingly, $\eta$ represents an additional dephasing rate beyond the rate induced by the gain saturation.
If $\eta=0$ ($\eta\neq 0$) the system is a standard (generalized) SLLM.

The Lindblad form of the SLLM master equation is well-defined in
the weak-gain saturation (WGS) regime, valid under the condition
\begin{equation}\label{Eq:Weak-gain-saturation}
    B\ll 1, \, A-\Gamma \ll 2 A, \quad {\rm and} \quad \,
    {B}\expec{\hat{a}\hat{a}^\dagger}\ll {2A}.
\end{equation}
The WGS conditions are satisfied around the lasing
threshold~\cite{YamamotoBook}. Far away from the threshold,
the jump operator $\hat L_1$ in \eqref{Eq:L123} can be nonphysical, and a different master equation needs to be employed~\cite{ScullyLambBook,YamamotoBook}. 
Since the main aim of this paper is to investigate the nonequilibrium phase transitions of the laser at the threshold, we can safely use $\hat L_{1, \, 2, \, 3}$ in \eqref{Eq:L123} (see also Appendix~\ref{APP:WGS}).

\subsection{The lasing transition within the semiclassical approximation}

As discussed in Ref.~\cite{MingantiarXiv20}, for this $U(1)$ model we can set $\omega=0$ via a reference frame change: from the laboratory one to the one rotating at the cavity frequency. 
This leaves the overall properties of the system unchanged.
 Within such a rotating frame, the semiclassical approximation (often---but not always---justified in the limit of intense fields)
 amounts to consider that the correlation functions $\expec{\hat{a}^{\dagger\, m}\hat{a}^{n}} = \expec{\hat{a}^{\dagger\, m}}\expec{\hat{a}^{n}}\to\alpha^{*m}\alpha^n$. Equivalently, the density matrix $\rhot$ of the system is, at all times, in the \textit{coherent state} $\rhot=\ket{\alpha}\bra{\alpha}$. Under this approximation, and working in the WGS regime of \eqref{Eq:Weak-gain-saturation}, one can compute the
mean-photon number in the steady state~\cite{OrszagBook,Gea1998}, obtaining either a trivial solution $\expec{\hat{a}^\dagger
\hat{a}}_{\rm ss}=n_{\rm ss}=0$ or
\begin{equation}\label{Eq:Semiclassical_solution} \expec{\hat{a}^\dagger \hat{a}}_{\rm ss}=n_{\rm ss}= |\alpha|_{\rm ss}^2 = \frac{A-\Gamma-\eta}{B}. \end{equation}
Stability analysis reveals that, for
$A>\Gamma+\eta$, the solution $n_{\rm ss}=0$ becomes unstable, and the
system passes nonanalytically from zero to a nonzero number of
photons. 

Thus, the system undergoes a \textit{second-order phase transition}, whose order parameter $\alpha=\expec{\hat{a}}_{\rm ss}\neq 0$ indicates the $U(1)$ SSB of the model (see the discussion in Sec.~\ref{Sec:Symmetry_Division}). 
Let us remark that the semiclassical equations of motion for the fields obtained from Eq.~(\ref{Eq:Lindblad1}) do \textit{not} include any contribution from quantum fluctuations.
We also note that a rigorous quantum solution for the steady state reveals that at the threshold $\expec{\hat{a}^\dagger
\hat{a}}_{\rm ss}\neq0$~\cite{ScullyLambBook,YamamotoBook}.

\subsection{The full quantum solution}
\label{Subsec:P-repre} 

Let us analytically proof that the semiclassical analysis is wrong in the case $\eta\neq 0$, thus demonstrating the necessity to go beyond the semiclassical approximation to correctly describe the lasing transition of the SLLM.

The Glauber-Sudarshan $P$-representation is a mapping of the
density matrix from the bosonic Hilbert space to the complex phase
space $\mathbb{C}$ via a real-valued function $P(\alpha)$, which
reads
\begin{equation}
    \rhot= \int \de^{2} \alpha \; P(\alpha, t) \ket{\alpha}\bra{\alpha},
\end{equation}
where $\ket{\alpha}$ is a coherent state
\cite{Carmichael_BOOK_1,Gardiner_BOOK_Quantum,Walls_BOOK_quantum}.
In this formalism, the matrix equation for $\rhot$ is transformed
into a differential equation for the quasiprobability function
$P(\alpha, t)$~\cite{Vogel_PRA_89_quasiprobability,Lu1989}
by repeatedly applying the rules:
\begin{eqnarray} \left\{
\begin{array}{l} \rhot\hat{a}\\
\hat{a}\rhot
\end{array}
\right\} &\mapsto& \left( \alpha -\frac{1\pm 1}{2}
\frac{\partial}{\partial \alpha^*}\right) P(\alpha, t),
\nonumber\\
\left\{ \begin{array}{l}
\hat{a}^\dagger\rhot\\
\rhot\hat{a}^\dagger
\end{array}
\right\} &\mapsto& \left( \alpha^* -\frac{1\pm 1}{2}
\frac{\partial}{\partial \alpha}\right) P(\alpha, t). 
\end{eqnarray}
Under the weak-gain-saturation condition in \eqref{Eq:Weak-gain-saturation}, 
\eqref{Eq:L123} becomes
\begin{widetext}
\begin{equation}\label{Eq:P-representation-evolution}
\frac{\partial P(\alpha,t)}{\partial
t}=-\frac{1}{2}\left\{\left[\frac{\partial}{\partial\alpha}\left(A-\Gamma-\beta-B|\alpha|^2\right)\alpha+
{\beta}\frac{\partial^2}{\partial\alpha^2}\alpha^2\right]P(\alpha,t)+\text{c.c.}\right\}+\frac{\partial^2}{\partial\alpha\partial\alpha^*}\left(A+\beta|\alpha|^2\right)P(\alpha,t).
\end{equation}
In the steady state $\partial_t P_{\rm ss}(\alpha)=0$, the $P$
function depends only on the field intensity $|\alpha|^2$. By
rewriting Eq.~(\ref{Eq:P-representation-evolution}) in the polar
coordinates $(r,\phi)$, one can separate $P(\alpha, t)$ into
its radial [$P(r,t)$] and angular [$P(\phi,t)$] components. This
is also true for the steady state, and $P_{\rm ss}(r,\phi)=P_{\rm
ss}(r) P_{\rm ss}(\phi)$. At the steady state, $P_{\rm ss}(\phi) =
1/(2 \pi)$, while
\begin{equation}\label{Eq:Psseq}
-\frac{1}{2}\frac{1}{r}\frac{\partial}{\partial
r}\left[r^2\left(A-\Gamma-B r^2\right)P_{\rm
ss}(r)\right]+\frac{A}{4}\left(\frac{\partial^2}{\partial
r^2}+\frac{1}{r}\frac{\partial}{\partial
r}\right)P_{\rm ss}(r)=0.
\end{equation}
\end{widetext}
By solving Eq.~(\ref{Eq:Psseq}), one obtains
\begin{equation}\label{Eq:Pss}
P_{\rm ss}(\alpha)=P_{\rm ss}(\phi) P_{\rm
ss}(r)=\frac{\exp\left[\frac{r^2}{A}\left(A-\Gamma-B{r^2}\right)\right]}{2
\pi \mathcal{N}},
\end{equation}
where $\mathcal{N}$ is a normalization constant. 

In striking contrast to the semiclassical approach, the parameter $\eta$ does not appear in the steady-state solution (\ref{Eq:Semiclassical_solution}). 
As such, the evident contradiction between the photon number obtained using the semiclassical approximation in \eqref{Eq:Semiclassical_solution} (depending on $\eta$) and the $P$-representation in the weak-gain-saturation regime (which does not depend on $\eta$) cannot be reconciled, even in the thermodynamic limit. In other words, the effect of the photon dephasing is overrated in determining the intensity of the field in the semiclassical approach, and a (superposition of) coherent state(s), with the amplitude in Eq.~(\ref{Eq:Semiclassical_solution}), can never be the steady state of the system for $\eta \neq 0$. We provide a detailed description of this semiclassical approximation failure in Appendix~\ref{Sec:Quantum_Fluctuations}.

\subsubsection{Diffusion coefficient and the loss of coherence}

Although $\eta$ plays no role in the determination of the steady state, it strongly affects the dynamics of the system. 
To prove the last affirmation, we notice that the loss of
coherence is encoded in the evolution of $P(\phi, t)$. We can
estimate such time by considering an initial state $P(r, \phi,
t=0) = P_{\rm ss}(r) \delta(\phi - \phi_0)$. Accordingly, we have
[c.f. the right-hand side of the  \eqref{Eq:P-representation-evolution}]:
\begin{equation}\label{Diffusion_eq}
    \frac{{\rm d}}{{\rm d}t}P(\phi, t)=\frac{1}{4}\left(\frac{A}{\expec{\hat n}_{\rm ss}}+\beta\right)\frac{\partial^2}{{\partial}\phi^2}P(\phi, t).
\end{equation}
This is just a diffusion equation with a diffusion
coefficient
\begin{equation}\label{D}
D=\frac{A}{2\expec{\hat n}_{\rm ss}}+\frac{\beta}{2}.
\end{equation}
The diffusion coefficient $D$ determines the spectral line of a
laser. In the standard Scully-Lamb laser theory, the parameter
$\eta = 0$, $B/A \ll 1$, and the laser line tends to zero
with increasing photon number $A/\expec{\hat n}_{\rm ss}\to0$.
Thus, as expected, a state will maintain its coherence even for infinitely large times $t\to \infty$.

On the contrary, in the generalized SLLM, the laser
linewidth remains finite, since $D=\beta/2 \simeq  \eta/2$ in
the large-gain limit. 
Thus the $P$ function already hints at the
fact that the system cannot break the $U(1)$ symmetry, since it cannot retain a phase.
That is, the generalized SLLM,  in the presence of $\eta \neq 0$, cannot be characterized by a $U(1)$ SSB. 

We can already
see the profound and nontrivial interplay between the dynamics of
the system, the symmetry breaking, and the emergence of a phase
transition.

\subsection{The Landau theory}
\label{Sec:LandaU_Theory}
One can try to go beyond the semiclassical analysis of the phase transition by using the well-known \textit{Landau theory} of phase transitions, assuming an analogy with thermal
phase transitions in equilibrium systems. 
This approach seems particularly effective in the standard SLLM (for which $\eta\to0$). Indeed, the exponent of $P_{\rm ss}$
in Eq.~(\ref{Eq:Pss}) can be directly associated with the ``free energy'' $G$ of the system~\cite{DeGiorgio1970}.
Remarkably,
the relation ${\dot{E}}\equiv-\frac{\partial G}{{\partial}{E}}$
can be derived, where $ E\equiv\alpha_{\rm ss}+\alpha_{\rm ss}^*$ is the electric field obtained via the semiclassical approximation in \eqref{Eq:Semiclassical_solution}.
From the Landau theory one deduces the presence of
a \textit{second-order phase transition}, signalled by a
nonanalytical change in the intensity of the electric field
(i.e., the photon number). This is associated with the $U(1)$ SSB in the system.

Within the generalized SSLM ($\eta\neq 0$), no conclusions
about such a correspondence between the $P$-function and the
corresponding $G$ can be made. Even
if blindly assuming that the exponent of $P_{\rm ss}$ still
accounts for the free energy, one would erroneously conclude that still
the phase transition, associated with the SSB of $U(1)$, is taking place~\cite{DeGiorgio1970}.

\section{Dissipative Phase Transition of the Scully-Lamb model}
\label{Sec:Intro_Spectral_collapse}

A dissipative phase transition is a discontinuous change in the steady state $\sss$ as a function of a single
parameter~\cite{MingantPRA18_Spectral,KesslerPRA12}. In the following analysis, the gain $A$ in \eqref{Eq:L123} plays such a role (see, e.g., Fig.~\ref{fig:liouvillian_gap}), and the definition for the second-order dissipative phase transition reads
\begin{equation}\label{Eq:definition_PT}
    \lim_{A \to A_c} \frac{\partial^{2}}{\partial{A}^{2}} \sss(A)  \to \infty,
\end{equation}
where $A_c$ is the critical point.

This nonanalyticity occurs in the thermodynamic limit. However,
the effects of an emerging criticality can also be witnessed in finite-size systems. Indeed, there is a profound connection
between the dynamical properties of a dissipative system and the
emergence of a phase transition, as proved in
Refs.~\cite{KesslerPRA12,MingantPRA18_Spectral} and experimentally
demonstrated
in Refs.~\cite{RodriguezPRL17,FitzpatrickPRX17,FinkNatPhys18}.
Criticality is accompanied by the emergence of the so-called
\textit{critical slowing down}, i.e., the appearance of
infinitely long timescales in the system dynamics. 

To properly characterize the emergence of a critical timescale, we
resort to the Liouvillian spectrum. Mathematically, we can define
the eigenvalues $\lambda_i$ (representing the time scale of the
problem) and eigenvectors $\eig{i}$ (encoding the states explored
along the dynamics) of $\LL$ via
\begin{equation}\label{Eq:Definition_spectrum}
\LL \eig{i}=\lambda_i \eig{i}.
\end{equation}
While $\re{(\lambda_i)}$ indicates a decay time towards the steady
state, $\im{(\lambda_i)}$ encodes the frequency of the associated
oscillations.

The Liouvillian is a superoperator, which acts on operators to generate new operators, in the same way in which an operator acts on a vector to generate a new vector.
We use the symbol $\bigcdot$ to indicate the density matrix placeholder. For example, $- i [\hat{H}, \bigcdot] \rhot= - i [\hat{H}, \rhot]$.
Since the Liouvillian is a linear superoperator [i.e., $\LL (a \hat{A} + b \hat{B})= a \LL \hat{A} +b \LL \hat{B} $], $\LL$ can be represented in a matrix form.
A detailed discussion of superoperator properties can be found in, e.g., Refs~\cite{Carmichael_BOOK_2,LidarLectureNotes,MingantiPRA19}.

\subsection{The thermodynamic limit}

Phase transitions and nonanaliticity can only emerge in the thermodynamic limit. 
Thus, we need to introduce a well-defined thermodynamic limit to properly compare the Liouvillian results with those of the Landau theory and the semiclassical approach discussed above. 

In a lattice system with $L$ sites, this corresponds to increasing the size $L$ to infinity. In bosonic systems, one can exploit the infinite dimension of the Hilbert space to observe a nonanalytical change in the steady state, as discussed, e.g., in Refs.~\cite{CarmichaelPRX15,CasteelsPRA17,BartoloPRA16,MingantPRA18_Spectral,CurtisPRR21,MingantiarXiv20}.
To do so, one considers an appropriate rescaling of the parameters, so that the size of the Hilbert space increases, but the rescaled observables merge far from the critical point. 
We introduce an effective scaling parameter $N$ transforming the photon number $n \to N n=N(A-\Gamma)/B$ in \eqref{Eq:Semiclassical_solution}.
Hence, any transformation
\begin{equation}\label{Eq:Scaling}
\{A, B, \Gamma\} \to \{A N^{\mu}, B/N^{(1- \mu)}, \Gamma N^{\mu}
\}
\end{equation}
is valid for any real number $\mu$, and the thermodynamic limit is
reached by increasing $N$.
While the steady state is unchanged by different choices of $\mu$, the choice
$\mu=0$ ensures that the dissipation rate $\Gamma$ is kept constant (providing a natural timescale) for the problem. 

Considering the laser model of Ref.~\cite{YamamotoBook} and Appendix~\ref{Sec:Non_Lindblad_ME}, where the gain is caused by the injection of inverted atoms, the scaling parameter $N$ in \eqref{Eq:Scaling} corresponds to decreasing the light-matter coupling $g$, while increasing the pumping rate (i.e., without increasing the gain $A$). Thus, the parameter $N$ can be seen as the number of constantly pumped atoms required to compensate the decreasing light-matter interaction.
The thermodynamic limit $N\to \infty$ is well defined, since we are increasing the number of particles, while keeping constant the interaction energy density.

\begin{figure}
    \centering
    \includegraphics[width=0.49 \textwidth]{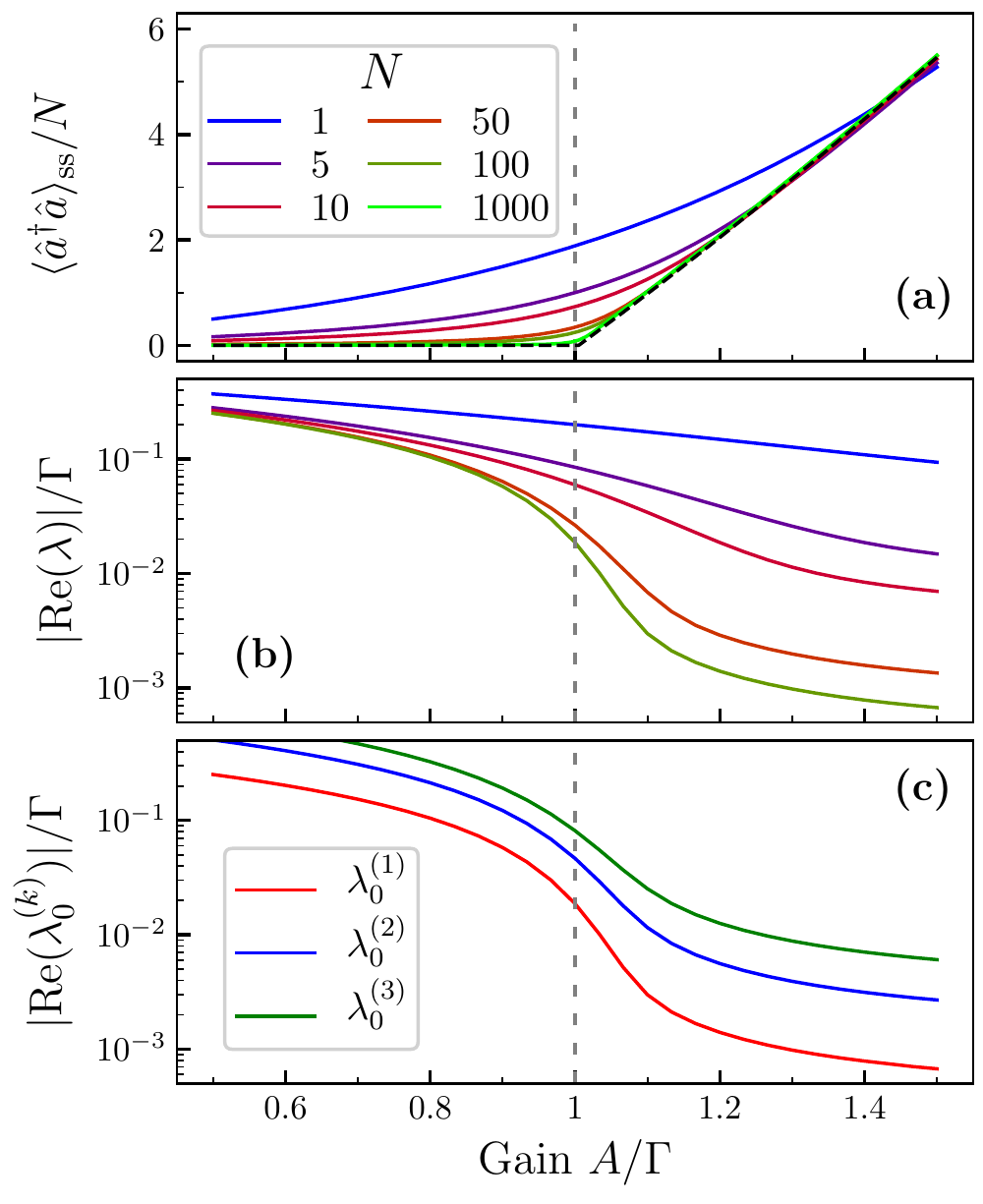}
    \caption{Demonstration of the $U(1)$ symmetry breaking of the standard SLLM ($\eta=0$) in the Liouvillian description.
    As a function of the incoherent drive strength $A/\Gamma$: (a) the rescaled number of photons $\expec{\hat{a}^\dagger \hat{a}}_{\rm ss}/N$, (b) the real part of the Liouvillian gap $\lambda$ [c.f. \eqref{Eq:Gap}], and (c) the real part of the eigenvalue closest to zero in each symmetry sector $\lambda_0^{(k)}$ for $k=1,\,2,\,3$. The black dashed lines in (a) represent the semiclassical coherent state approximation in \eqref{Eq:Semiclassical_solution}, and the corresponding critical point is marked by the vertical dashed gray line.
    Parameters: the gain saturation rate $B/\Gamma=10^{-1}/N$, $\omega/\Gamma=1$ (the plotted results are independent of $\omega$), and $\eta=0$. In (c), we set $N=100$.
    }
    \label{fig:liouvillian_gap}
\end{figure}

In Fig.~\ref{fig:liouvillian_gap}(a), we plot the photon number $\expec{\hat{a}^\dagger \hat{a}}/N$, obtained by numerically solving $\LL \sss =0$, for different values of $N$. The photon number becomes sharper and sharper with increasing $N$, converging towards the semiclassical result of
\eqref{Eq:Semiclassical_solution}. This confirms the presence of criticality in the thermodynamic limit, and corroborates the analysis of the scaling parameter $N$ in \eqref{Eq:Scaling}.

\subsection{Symmetry breaking}

\label{Sec:Symmetry_Division}

We can refine the spectral analysis in \eqref{Eq:Definition_spectrum} by introducing the symmetries of the Liouvillian. The SLLM is characterized by a $U(1)$ symmetry of the Lindblad master equation, because the transformation $\hat{a}
\to \hat{a} \cdot \exp(i\phi)$ leaves the Liouvillian unchanged for
any real number $\phi$. Thus, the superoperator
\begin{equation}\label{Eq:U_1_superoperator}
\mathcal{U}  = \exp\left(-i \phi \hat{a}^\dagger \hat{a} \right)\, \bigcdot \,\left(i
\phi \hat{a}^\dagger \hat{a}\right),
\end{equation}
commutes with the Liouvillian, i.e., $[\LL, \mathcal{U}]=0$. A Hamiltonian symmetry implies the presence of a conserved
quantity. This is not always the case for Liouvillian
symmetries~\cite{BucaNPJ2012,AlbertPRA14,BaumgartnerJPA08}.
However, all eigenmatrices of $\mathcal{L}$ must be eigenmatrices
of $\mathcal{U}$~\cite{BaumgartnerJPA08, MingantPRA18_Spectral,
AlbertPRA14,AlbertPRX16}; that is,
\begin{equation}\label{Eq:symmetry_on_eigenmatrices}
\mathcal{U} \eig{i} = u_k \eig{i} = \exp \left( i k \phi \right)\eig{i},
\end{equation} where $u_k$ is one of the eigenvalues of $\mathcal{U}$, and $k$ is an integer (see Appendix \ref{Sec:Eigenmatrices_Wigner} for more details).

We refer to the span of all the eigenmatrices characterized by the
same $u_k$ as a \emph{symmetry sector}. Each symmetry sector
is a part of the Liouvillian space which is not connected
to its other parts (sectors) by the Liouvillian
dynamics~\cite{MoodiePRA18, MingantiarXiv20,PalacinoPRR21}. In
other words, the Liouvillian can be decomposed into a direct sum
of superoperators $\LL_k$ acting on different symmetry sectors as
\begin{equation}
    \LL = \LL_{0} \oplus \LL_{1} \oplus \LL_{-1} \oplus \LL_{2} \oplus \LL_{-2} \oplus \LL_{3} \oplus \dots
\end{equation}
As such, we can relabel the eigenvalues and eigenmatrices as
$\lambda_i^{(k)}$ and $\eig{i}^{(k)}$, respectively, where ${k}$ tracks the
symmetry sector:
\begin{equation}
    \LL_{k} \eig{i}^{(k)}= \lambda_i^{(k)} \eig{i}^{(k)}.
\end{equation}
We order the eigenspectrum of each sector
according to
\begin{equation}
    \left|\Re{\lambda_0^{(k)}}\right|<\left|\Re{\lambda_1^{(k)}}\right|
<\dots <\left|\Re{\lambda_n^{(k)}}\right| < \dots
\end{equation}

Using this convention, the steady state is proportional to
$\eig{0}^{(0)}$, i.e., the eigenmatrix of the Liouvillian which
does not evolve, being associated with
$\lambda_0^{(0)}=0$~\cite{MingantPRA18_Spectral}. Different $u_k$'s
represent different upper and lower diagonals [see \eqref{Eq:condition_symmetry_U(1)}].

We define the \textit{Liouvillian gap} $\lambda$ as the nonzero
eigenvalue of the Liouvillian whose real part is the closest to
zero, i.e.,
\begin{equation}\label{Eq:Gap}
    \lambda= \lambda_j^{(k)} \mbox{ such that } 0<\left| \operatorname{Re} \left[\lambda_j^{(k)}\right] \right| \leq \left| \operatorname{Re} \left[\lambda_p^{(q)}\right]\right| \, \forall p, \, q.
\end{equation}
The inverse of $\lambda$ indicates the
slowest relaxation time of the system. The closure of the gap
($\lambda\to 0$) indicates a diverging timescale and, thus,
criticality.

Within this representation, we can write any initial density matrix as
\begin{equation}\label{Eq:Spectral_decomposition}
    \hat{\rho}(0) = \sum_{j, \, k} c_j \eig{j}^{(k)}, 
\end{equation}
for an appropriate choice of coefficients $c_j$, and, thus, we have
\begin{equation}\label{Eq:Spectral_decomposition_time}
    \rhot = \sum_{j, \, k} c_j \eig{j}^{(k)} \exp\left(\lambda_{j}^{(k)} t \right).
\end{equation}

\subsubsection{Spontaneous breaking of the $U(1)$ symmetry}

In a Liouvillian framework, the breaking of a $\mathcal{Z}_N$
symmetry takes place  when $(N-1)$
eigenvalues, each belonging to a different symmetry sector, become
zero: $\lambda_{0}^{(N-1)},\dots,\lambda_0^{(1)}=0$
(see Ref.~\cite{MingantPRA18_Spectral}). These eigenvalues must become
zero not only at the critical point, but also in the entire
broken-symmetry regime, ensuring that there exist several steady
states which are not eigenstates of the symmetry operator. Since $\mathcal{U}=\lim_{N \to \infty} \mathcal{Z}_N$, a SSB in the SLLM means that in each symmetry sector (except $u_0$) there is a vanishing eigenvalue. As suggested by the
semiclassical analysis, we should observe this feature 
when $A>A_c =\Gamma$ [c.f. \eqref{Eq:definition_PT}]. 

In Fig.~\ref{fig:liouvillian_gap}(b), we demonstrate that,
converging towards the thermodynamic limit $N\to \infty$, the
Liouvillian gap [defined in \eqref{Eq:Gap}] tends to zero for $A>A_c$. This indicates a symmetry breaking, as also confirmed for similar models in Refs.~\cite{TakemuraJOSAB21,Munozarxiv20}. 
To prove that the SSB is that of $U(1)$,  in Fig.~\ref{fig:liouvillian_gap}(c) we plot $\lambda_0^{(k)}$ for $k=1, \, 2, \,3$ for fixed $N=100$. In each sector we observe a similar ``closure'' of the gap, confirming the presence of $U(1)$ symmetry breaking. We verified that in all sectors (up to the numerical cutoff) the closure of the Liouvillian gap becomes more evident as we increase $N$.

Technically speaking, the SLLM lasing transition is a
\textit{time crystal}. The imaginary part of the plotted eigenvalues is never
zero ($\operatorname{Im}[\lambda_{0}^{(k)}]=k \omega $), and therefore undamped oscillations take place.
However, as proved in Ref.~\cite{MingantiarXiv20}, the imaginary part of $\lambda_{0}^{(k)}$ can be set to zero via an appropriate change of the reference frame, resulting in a genuine symmetry-broken phase.

\begin{figure*}
    \centering
    \includegraphics[width=0.98\textwidth]{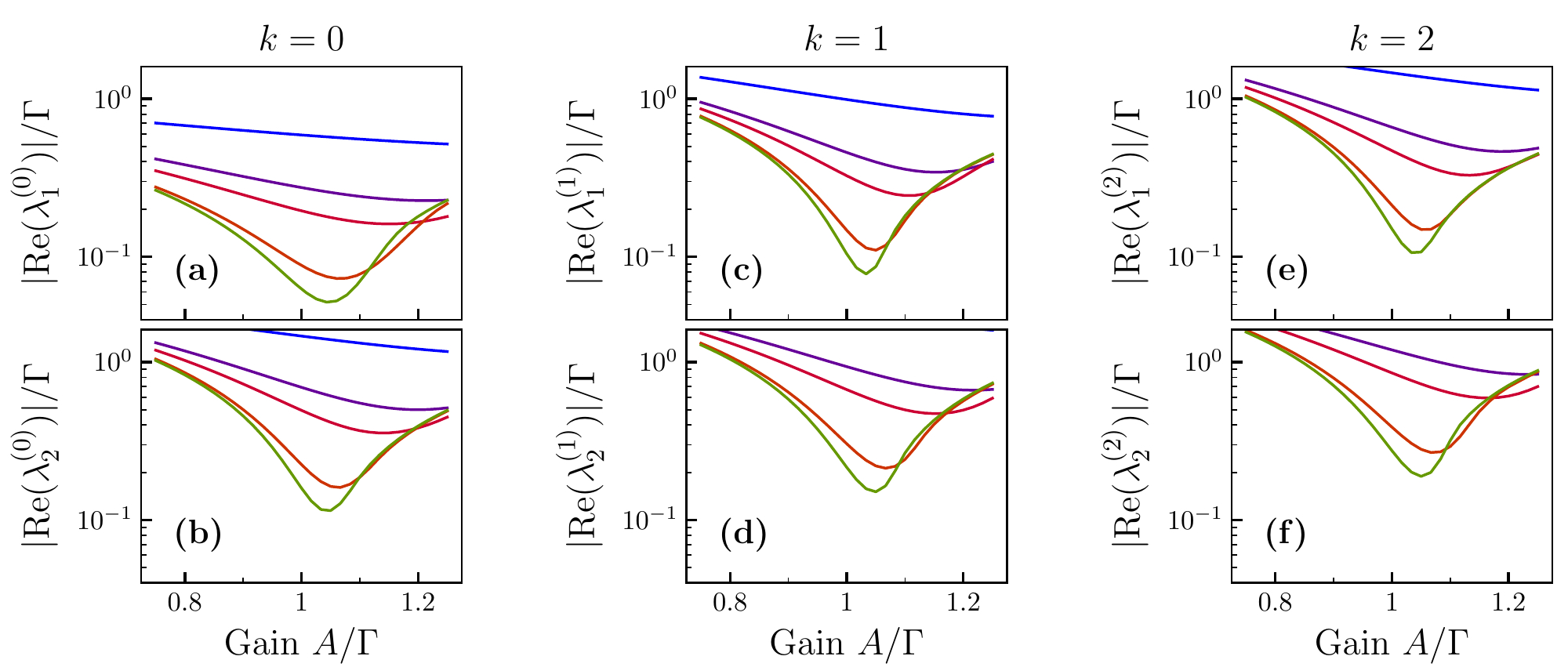}
    \caption{
Liouvillian spectral collapse of the standard Scully-Lamb laser model ($\eta=0$): in each sector $k$, several
eigenvalues, which are not associated with the SSB,
acquire a vanishing real part as we increase $N$. Here we plot the real
part of the second and third-largest eigenvalues, labeled as
$\lambda_1^{(k)}$ and $\lambda_2^{(k)}$, for the symmetry sectors
$k=0, \;1, \; 2$ and for different values of $N$. All the curves
converge towards $\lambda_j^{(k)}=0$ at the critical point
$A=\Gamma$ in the thermodynamics limit $N\to \infty$ (as can be
argued also from Fig.~\ref{fig:scaling}). Parameters and legend are the
same as in Fig.~\ref{fig:liouvillian_gap} (the results are
independent of $\omega$).}
    \label{fig:gap_in_zero}
\end{figure*}

\begin{figure}
    \centering
    \includegraphics[width=0.49 \textwidth]{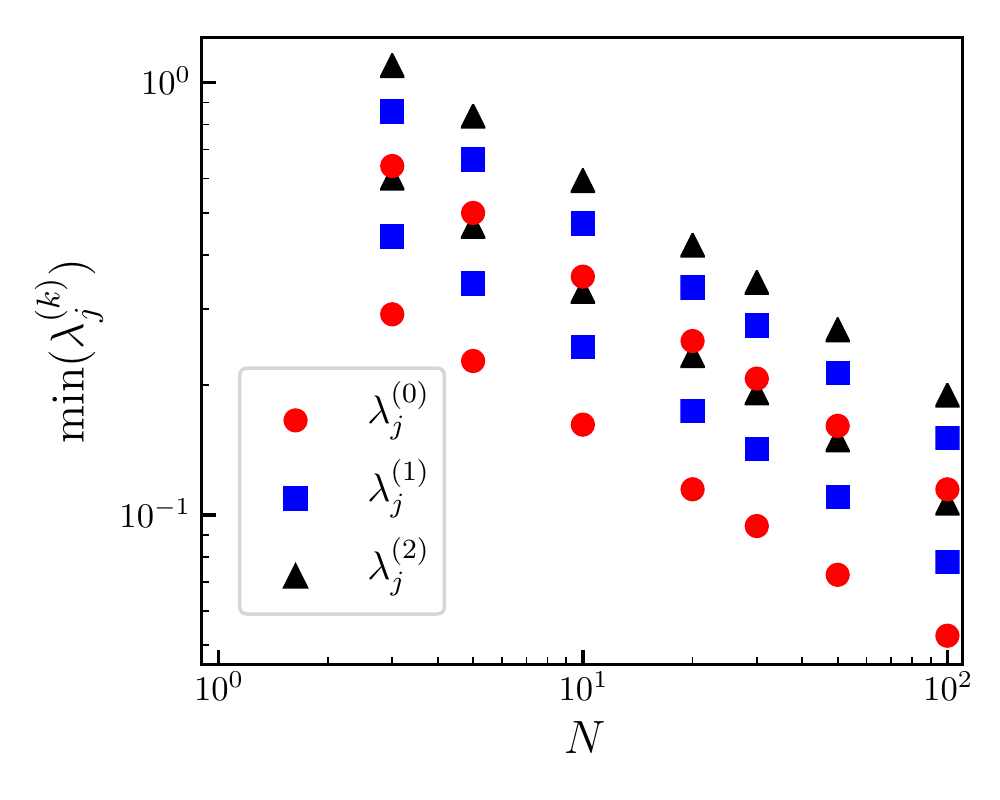}
    \caption{Minimum of the real part of the eigenvalues $\lambda_j^{(k)}$ versus the scaling parameter $N$,  for $j=1,\,2$ and $k=0,\,1,\, 2$, as also plotted in Fig.~\ref{fig:gap_in_zero}. While in Fig.~\ref{fig:gap_in_zero} we demonstrated that the position of the minimum tends to the critical point $A_c=\Gamma$, here we can clearly see that all the eigenvalues have the same convergence rate, further demonstrating the existence of a spectral collapse.
    Parameters are the same as in Fig.~\ref{fig:liouvillian_gap}.}
    \label{fig:scaling}
\end{figure}

\subsection{Spectral collapse in the standard Scully-Lamb laser model}
\label{Sec:Liouvillian_theory} 
For this standard SSLM ($\eta=0$) the results of the Landau theory
and those of the semiclassical approximation agree with the 
Liouvillian theory in predicting the $U(1)$ SSB ~\cite{DeGiorgio1970}.
However, we just analyzed $\lambda_0^{(k)}$, i.e., the slowest relaxation rate in each symmetry sector, ignoring the 
remainder of the spectrum. By analyzing $\lambda_j^{(k)}$ for $j>0$, we show that the SLLM
transition is characterized by a different kind of criticality,
which we call a \textit{Liouvillian spectral collapse}. 
For this standard SLLM, this does not mean that the SSB does not take place, but rather that the lasing threshold has a richer structure.

\label{Sec:Spectral_collapse}

In Fig.~\ref{fig:gap_in_zero}(a), we plot the smallest nonzero
eigenvalue in the sector $u_0$, namely, $\lambda^{(0)}_1$. The
inverse of $\lambda^{(0)}_1$ describes the slowest relaxation rate
of all those observables of the form $\expec{(\hat{a}^\dagger)^m
\hat{a}^m}$. 
Surprisingly, we observe that $\lambda^{(0)}_1$ becomes zero at the critical point. Note that the sector $u_0$ comprises the Liouvillian
eigenmatrices which are phase-independent (i.e., diagonal such as the steady state). 
Consequently, the SSB, which is associated with the development of a
well-defined phase, is not directly associated with the criticality in $u_0$. 
The absence of a critical timescale in $u_0$ has been observed in models with a $Z_2$ SSB, as can be inferred in, e.g., Refs.~\cite{MingantPRA18_Spectral,RotaNJP18}.
We stress that, to numerically distinguish $\lambda^{(0)}_1$ from all the other eigenvalues, which become zero at the transition, one
needs to properly divide the system in its symmetry sectors, as detailed in Sec.~\ref{Sec:Symmetry_Division}.

Remarkably, this point-like closure of the gap occurs for all the symmetry sectors for the same critical value of $A_c=\Gamma$ [we
plot those of $\lambda^{(1)}_1$ and $\lambda^{(2)}_1$ in
Figs.~\ref{fig:gap_in_zero}(b) and \ref{fig:gap_in_zero}(c)], and for multiple
eigenvalues in this sector [$\lambda^{(k)}_2$ in Figs.~\ref{fig:gap_in_zero}(d),
\ref{fig:gap_in_zero}(e), and \ref{fig:gap_in_zero}(f)].
We numerically tested that, by increasing $N$, more and more eigenvalues collapse, in their real
part, towards zero.

One might think that a possible (although involved) explanation of
this phenomenon is that the real part of each eigenvalue
converges to zero independently of the others, so that some
eigenvalues are ``more critical'' than the others.
However, this conjecture is false. In Fig.~\ref{fig:scaling}, we plot the
minimum of the eigenvalues $\lambda_j^{(k)}$ across the transition
for those eigenvalues which present a spectral collapse (i.e.,
$\lambda_j^{(k)}$ for $j>0$). Obviously, the convergence rate of
all these eigenvalues is identical, meaning that they are all
resulting from the same phenomenon.

In other words, we demonstrated that the lasing threshold according to the SLLM
corresponds to an infinite degeneracy of \textit{nondecaying states} at
the critical point. 
Mathematically speaking, there is a \textit{diabolic point of an infinite order in each of the infinitely many symmetry sectors of the Liouvillian}. We are sure that the criticality is not associated
with exceptional points since the degeneracies emerging in the thermodynamic limit, where
$\re{[\lambda_j^{(k)}]}=0$ prevents the occurrence of exceptional points~\cite{MingantiPRA19}.

We refer to the emergence of infinitely degenerate manifolds as a \textit{Liouvillian spectral collapse}, in analogy to the
definition applied to the two-photon Rabi model~\cite{FelicettiPRA15}. Notice that, while the real part of
$\lambda^{(k)}_j \to 0$, the imaginary part is always fixed by the symmetry sector, since $\im{\left[\lambda^{(k)}_j\right]}=k
\omega$.

Let us remark that this degeneracy does not imply that all the processes are infinitely long lived. 
Indeed, there is a part of the spectrum which never collapses to zero. For
instance, at the critical point, a state initialized in a Fock state $\ket{n}\bra{n}$ rapidly
changes its parity, defined by $\hat{P}=\exp{(i \pi
\hat{a}^\dagger \hat{a})}$, on a timescale which is always of the
order of $1/ (\Gamma n)$.

This nonanalytical change occurs at the same point as
the $U(1)$ SSB. As we discuss in
more detail in the next section, the spectral collapse
is a fundamental ingredient of criticality, and the SSB is rather a byproduct of this process.

\section{Witnessing and characterizing the Liouvillian spectral collapse}
\label{Sec:Decoherence}

\begin{figure}
    \centering
    \includegraphics[width=0.49 \textwidth]{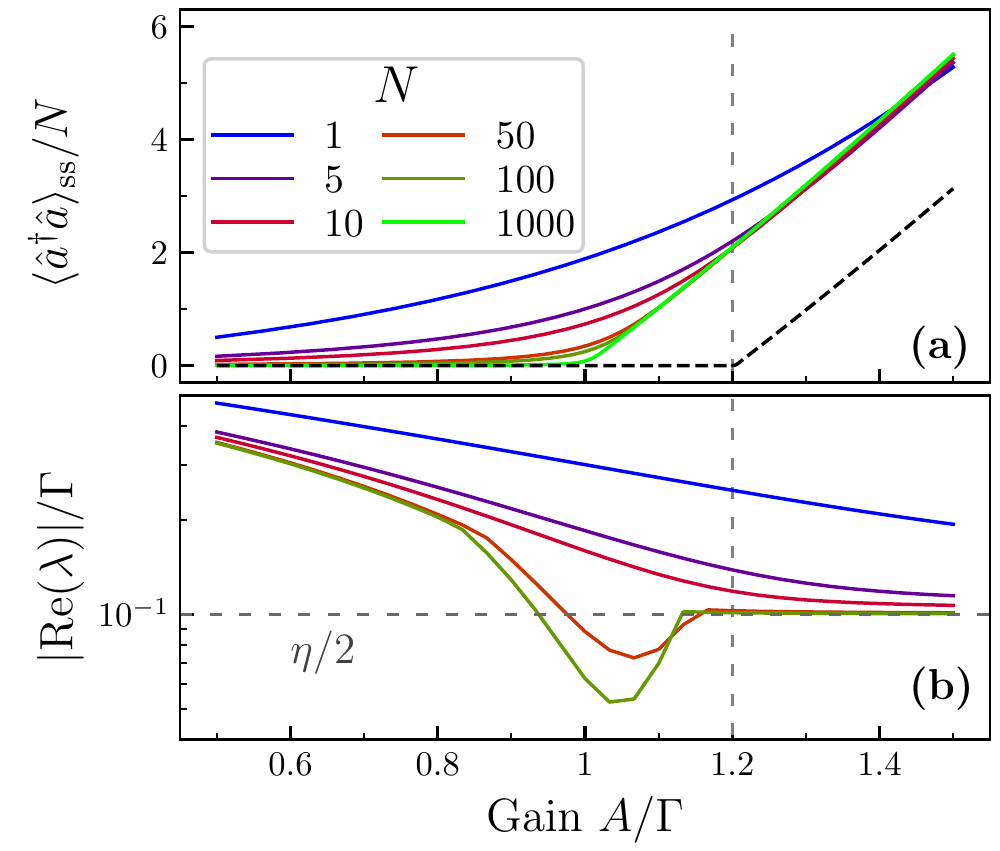}
    \caption{Criticality in the SLLM in the presence of additional decoherence $\eta  $. (a) Rescaled number of photons $\expec{\hat{a}^\dagger \hat{a}}_{\rm ss}/N$ versus the gain $A/\Gamma$. The black dashed lines represents the result of the semiclassical approximation, while the vertical dashed gray line indicates the semiclassical critical point. (b) In units of the damping rate $\Gamma$, Liouvillian gap $\lambda$ [defined in \eqref{Eq:Gap}] versus $A/\Gamma$. In this case, the gap closes at one point and does not indicate symmetry breaking. The gray horizontal dashed line is a guideline indicating the value $\eta/2$. Parameters:  saturation rate $B/ \Gamma=10^{-1}/N$, and $\eta=\Gamma/5$ (the results are independent of $\omega$).}
    \label{fig:num_phot_decoherence}
\end{figure}

As indicated by the $P$-representation in Sec.~\ref{Subsec:P-repre}, the steady state is unaffected by $\eta$.
On the contrary, the prediction of the semiclassical analysis in \eqref{Eq:Semiclassical_solution} is
that the phase transition point is shifted by $\eta$.
Indeed, $\eta$ cannot affect the dynamics of the steady-state symmetry sector, as it was recently demonstrated in Ref.~\cite{Minganti2021continuous}.
For this reason, by keeping $\eta$ constant even in the thermodynamic limit, it is possible to witness a second-order phase transition, even if we prevent SSB.

In Fig.~\ref{fig:num_phot_decoherence}(a), we show the validity of \eqref{Eq:Pss} and confirm the validity of the results in Ref.~\cite{Minganti2021continuous} because, despite an additional dephasing, the photon number is identical to that of the SLLM for $ \eta =0$ [c.f. Fig.~\ref{fig:liouvillian_gap}(a)]. 
Interestingly, \textit{the semiclassical coherent-state approach completely misses the onset of criticality}.
A detailed discussion of why the semiclassical approximation fails is given in Appendix~\ref{Sec:Quantum_Fluctuations}.
We also stress that a similar ``wrong'' prediction of SSB by a semiclassical analysis has been pointed out in Ref.~\cite{HuberPRA20} for a spin model, where the quantum solution shows a second-order dissipative phase transition without SSB.

As discussed in Sec.~\ref{Sec:LandaU_Theory}, a thermodynamic free energy, based on the $P$-function, would instead predict the presence of SSB. 
We conclude that such a prediction is wrong, and we cannot use the $P$-function as a thermodynamic potential within the Landau theory.

Notice that, in agreement with Ref.~\cite{Minganti2021continuous}, and as one can see from Fig.~\ref{fig:num_phot_decoherence}(b), the nonzero $\eta$ precludes any SSB (after the transition, there are no multiple steady states). 
However, the Liouvillian gap shows all the signs of the gap closure associated with the spectral collapse, as shown in Fig.~\ref{fig:gap_in_zero}.
Indeed, the spectral collapse takes place in the $u_0$ symmetry sector, i.e., those states which are unaffected by $\eta$, because they are symmetric under $U(1)$ action (see Appendix~\ref{Sec:Eigenmatrices_Wigner}).
For this reason, using the generalized SSLM, we can distinguish the effects induced by the spectral collapse from those
 depending on the SSB.

We also confirm the predictions of \eqref{D}. Indeed, after the phase transition occurs, $ \eta$ prevents the emergence of a symmetry-broken phase, and the Liouvillian gap for a large-enough $N$ is bounded by $ \eta/2$.
We conclude that the dephasing destroys the coherence of a laser state and no SSB is taking place because the retaining of a phase is the characteristic of the $U(1)$ SSB.
Further numerical simulations confirm that, for all values of $A$, an initial state with a nonzero phase rapidly looses its phase (on a timescale of the order of $\eta/2$).

\subsection{Anomalous multistability and dynamical hysteresis}
\label{Sec:hysteresis}

The presence of the Liouvillian spectral collapse and the degeneracy of eigenvalues is not just a theoretical artefact, because the existence of nondecaying processes at the critical
point has profound consequences in the physics of the SLLM which can be experimentally observed.

The presence of symmetry sectors of the Liouvillian affects also the dynamics of certain classes of operators. For instance, the evolution of $\hat{a}^\dagger \hat{a}$ depends only (explicitly or implicitly) on that of $\left(\hat{a}^\dagger\right)^m \hat{a}^m$. 
This correspondence between operators and density matrices becomes particularly useful when using the spectral decomposition in \eqref{Eq:Spectral_decomposition}.
For instance, since $\mathcal{U} (\hat{a}^\dagger \hat{a}) = \hat{a}^\dagger \hat{a}$ one can easily prove that
\begin{equation}
    \operatorname{Tr}\left[\eig{j}^{(k\neq 0)}  \hat{a}^\dagger \hat{a}\right]=0.
\end{equation}
That is, only the eigenmatrices of the symmetry sector $u_0$ have nonzero photon number expectation value.

\begin{figure}
    \centering
    \includegraphics[width=0.49\textwidth]{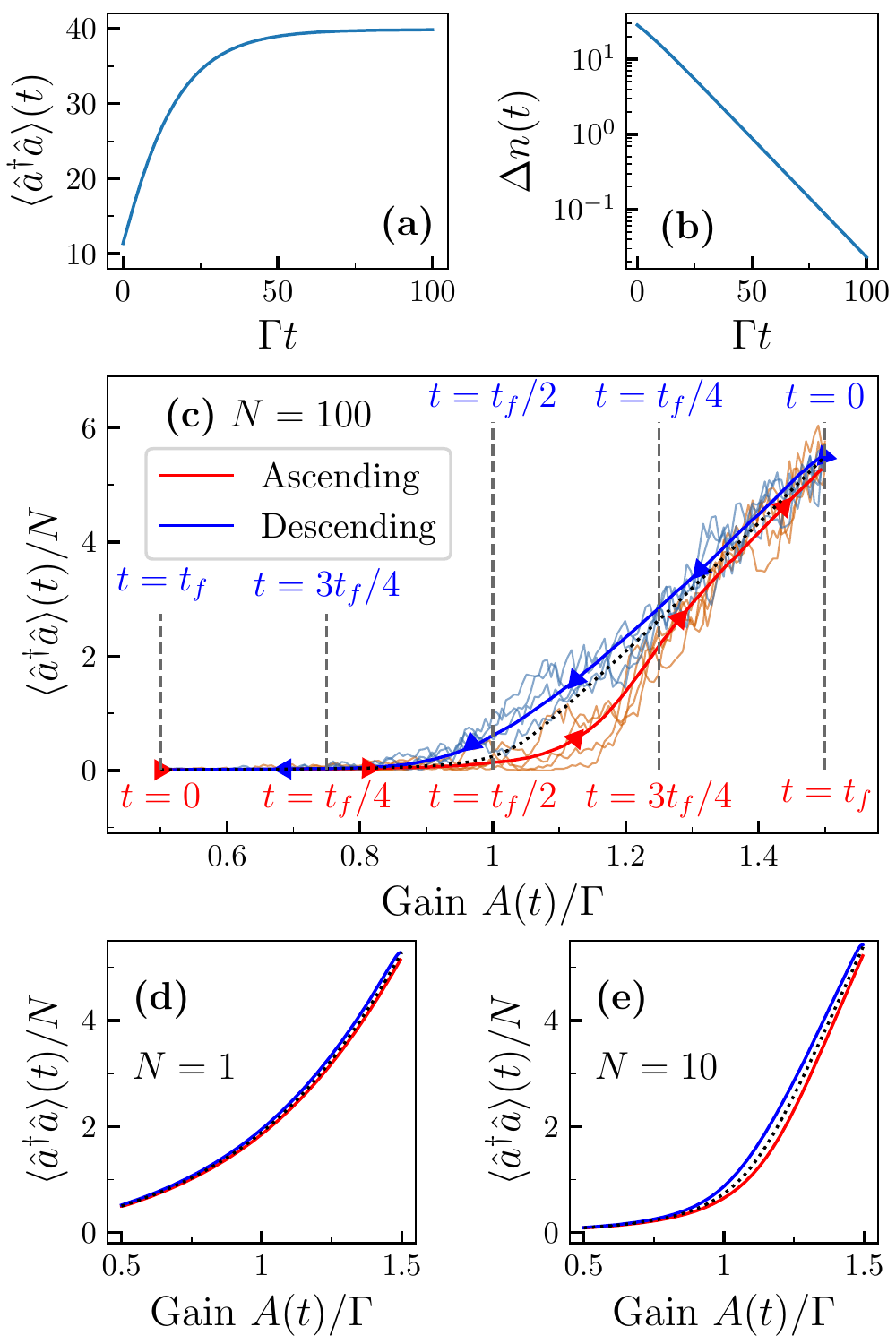}
\caption{Critical slowing down and dynamical hysteresis associated
with the Liouvillian spectral collapse.  Dynamics of (a) the photon number and
(b) the difference between the photon number in the steady state
and that at time $t$: $\Delta n(t) = \expec{\hat{a}^\dagger
\hat{a}}_{\rm ss} - \expec{\hat{a}^\dagger \hat{a}}(t)$. In (a)
and (b) we set $N=100$ and $A=1.05 \Gamma$ (ensuring that the gap in
the $u_0$ sector is minimal, c.f. Fig.~\ref{fig:gap_in_zero}), and
we initialize the system in a state with fewer photons than the those in the
steady state. [(c)-(e)] Hysteresis cycles in the photon number for an
ascending [red, \eqref{Eq:Ramp_up}] and descending [blue,
\eqref{Eq:Ramp_down}] dynamics for different values of $N$: (c)
 $N=100$, (d)  $N=1$, and (e)  $N=10$. The black dotted
curves are the expectation values of the number of photons in the
steady state. The thin curves in (c) represent single quantum
trajectories~\cite{MolmerJOSAB93,CarmichaelPRL93}. The initial
states have been chosen to be the steady state of $A(t=0)$. The final time is $t_{\rm f}=200 \Gamma$.
Parameters: the saturation rate $B/\Gamma=10^{-1}/N $ (the results are
independent of $\omega$ and $\eta$).}
    \label{fig:hysteresis}
\end{figure}

Therefore, the spectral collapse can be revealed by the photon-number evolution, since one is monitoring only the $u_0$ sector, i.e., the symmetric one which cannot experience criticality due to the $U(1)$ SSB [see Appendix~\ref{Sec:Eigenmatrices_Wigner} and Fig.~\ref{fig:wigner_eigenstates}(a)]. 
At the spectral collapse, the mean photon number $\expec{\hat{a}^\dagger \hat{a}}(t)$ \textit{must experience a critical slowing down}~\cite{LandaPRL20,LandaPRB20,MacieszczakPRL16}. Although the spectral collapse can be witnessed only for $N\to \infty$, a \textit{dynamical hysteresis} characterizes finite-size systems and can reveal an emerging criticality~\cite{RodriguezPRL17}.

To witness a dynamical hysteresis, one can slowly (with respect to $\Gamma$) change the value of $A$  [i.e., $\partial_t A(t) \ll \Gamma$]. Far from the critical point, we
expect the evolution to be adiabatic, since
$\lambda^{(0)}_1/\Gamma \simeq \mathcal{O}(1)$. Therefore,
$\expec{\hat{a}^\dagger \hat{a}}$ almost coincides with the
steady-state expectation value. However, around the critical
point, where $\lambda^{(0)}_1\ll \Gamma$, the evolution becomes
non-adiabatic. In other words, $\expec{\hat{a}^\dagger \hat{a}}$
retains the memory of its previous values.

We analyze the emergent critical slowing down and the dynamical hysteresis of the SLLM
lasing transition in Fig.~\ref{fig:hysteresis}. In
Figs.~\ref{fig:hysteresis}(a)~and~\ref{fig:hysteresis}(b), we
initialize the system to be close to the vacuum, and we observe
its evolution at the critical point.
Figure~\ref{fig:hysteresis}(a) shows that the time needed to reach
the steady-state photon number is much longer that $\Gamma$, which
represents the timescale far from the critical point in the
unbroken symmetry phase. Figure~\ref{fig:hysteresis}(b) shows
that, apart from the initial transient dynamics, the decay towards the
steady-state is purely exponential. An exponential fit of the
curve retrieves $\lambda^{(0)}_1$. Notice that, for finite
$N$, it holds $|\lambda_1^{(0)}|< |\lambda_2^{(0)}|$, and, therefore, the
long-time dynamics is  characterized solely by $\lambda_1^{(0)}$.

In Fig.~\ref{fig:hysteresis}(c), we show the presence of the dynamical
hysteresis for $N=100$. The red curve represents the master
equation evolution for a linear ramping of the parameter $A$, reading
\begin{equation}\label{Eq:Ramp_up}
A_{\uparrow}(t)=\frac{\Gamma}{2} + \frac{\Gamma t}{t_{\rm f}},
\end{equation}
for $t_{\rm f}=200 \Gamma$. Similarly, the blue curve represents the
inverse descending process
\begin{equation}\label{Eq:Ramp_down}
A_{\downarrow}(t)=\frac{3\Gamma}{2} - \frac{\Gamma t}{ t_{\rm f}},
\end{equation}
for the same $t_{\rm f}=200 \Gamma$. The thin superimposed curves represent the
results of a single quantum trajectory (i.e., reproducing the
result of a single ideal
experiment~\cite{DaleyAdvancesinPhysics2014}, see also Sec.~\ref{Sec:Novelty}). Far from the
critical point, i.e. $A \ll \Gamma$ or $A \gg \Gamma$, the
evolution is adiabatic, and the two curves overlap with that for the steady
state (black dotted curve). Instead, around the critical value,
the two dynamics are significantly separated, proving the presence of a
dynamical hysteresis.

In Figs.~\ref{fig:hysteresis}(d)~and~\ref{fig:hysteresis}(e), we show the results of
the same simulation but for $N=1$ and $N=10$. While for $N=1$
there are no signs of the hysteresis, for $N=10$ we observe some
minimal differences between the ascending and descending curves
around the critical regime. Comparing
Figs.~\ref{fig:hysteresis}(c),~\ref{fig:hysteresis}(d),~and~\ref{fig:hysteresis}(e)
we conclude that the hysteresis becomes more and more pronounced
with increasing $N$, confirming the results of
Fig.~\ref{fig:gap_in_zero}. Hence,
Figs.~\ref{fig:gap_in_zero}(a)~and~\ref{fig:gap_in_zero}(b), describe the emergence of a nonsymmetry breaking critical timescale.

Importantly, these results are independent of the additional dephasing in \eqref{Eq:L123} (all the plots in Fig.~\ref{fig:hysteresis} perfectly overlap).
Indeed, one can demonstrate that the $u_0$ symmetry sector is unaffected by the value of $\beta$.
Such a dynamical hysteresis, together with the nonanaliticity of the photon number in the cavity, could constitute an experimentally realizable test of the spectral collapse and of the presence of a second-order phase transition with or without $U(1)$ symmetry breaking.

We also stress that the presence of a dynamical hysteresis is not an artefact of the Liouvillian formulation of the SLLM, but similar results, based on the original model introduced by Scully and Lamb in Ref.~\cite{Scully1967}, have been obtained in Ref.~\cite{WangPRA73}.
Nevertheless, using the Liouvillian framework, we can assign a clear physical meaning to these phenomena.

\label{Sec:Characterization}

\subsection{Quantum fluctuations at the spectral collapse}
\label{Sec:Novelty}

\begin{figure}
    \centering
    \includegraphics[width=0.49 \textwidth]{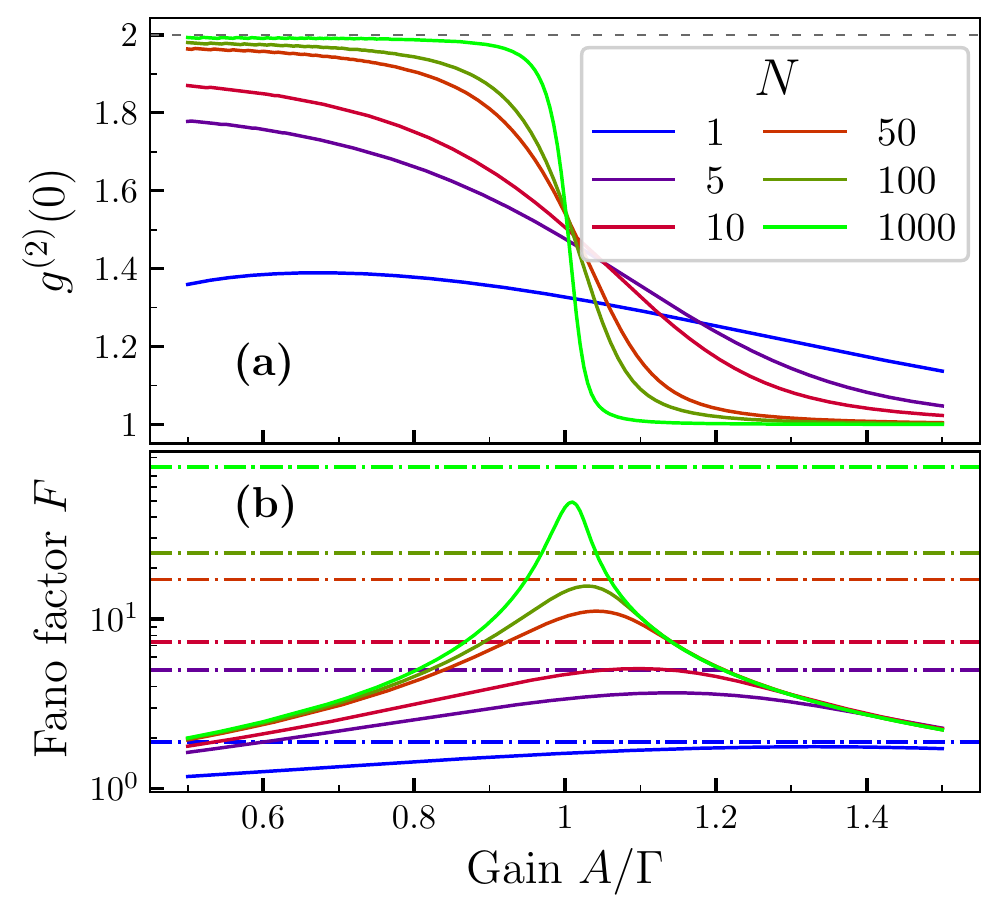}
    \caption{Effect of quantum noise in the SLLM at the critical point.
    (a) The second-order correlation function $g^{(2)}(0)$ (the gray-dashed line indicates the value 2), and
    (b) the Fano factor, defined in \eqref{Eq:Fano}, as a function of $A/\Gamma$. The dashed-dotted lines indicate $1+\langle\hat n\rangle$ at the critical point $A=\Gamma$, i.e., the value where light is classified as chaotic.
    Parameters are the same as in Fig.~\ref{fig:liouvillian_gap}.}
    \label{fig:correlations}
\end{figure}

To understand the cause of multistability at the spectral collapse, let us show that correlations play a fundamental role in the SLLM criticality.
First, let us consider the second-order single-time correlation function
defined as
\begin{equation}
    g^{(2)}(0)=\frac{\expec{\hat{a}^{\dagger\, 2} \hat{a}^2}_{\rm ss}}{\expec{\hat{a}^{\dagger} \hat{a}}^2_{\rm ss}}.
\end{equation}
We plot the results for $g^{(2)}(0)$ in
Fig.~\ref{fig:correlations}(a). We see that, at the critical point,
where the phase transition takes place, 
the value of $g^{(2)}(0)$ abruptly passes from 2 to 1. Remarkably, at the critical point, $g^{(2)}(0)\neq
1$, indicating that the system is not a coherent state.

To better visualize the incoherent nature of light at the transition, we use the Fano factor $F$, i.e.,
\begin{equation}\label{Eq:Fano}
    F = \frac{\langle(\Delta \hat n)^2\rangle}{\langle\hat n\rangle}
     = \langle\hat n\rangle [g^{(2)}(0)-1]+1,
\end{equation}
where $\Delta \hat{n}= \hat{a}^{\dagger} \hat{a} -
\expec{\hat{a}^{\dagger} \hat{a}}$. A coherent (Poissonian)
state is described by $F=g^{(2)}(0)=1$. If
$F<1$ and $g^{(2)}(0)<1$ ($F>1$ and $g^{(2)}(0)>1$) then it is
called sub-Poissonian (super-Poissonian), also referred to as
single-time photon antibunching (bunching). For a thermal chaotic
state, one has $g^{(2)}(0)=2$ and $F=1+\langle\hat n\rangle$.
Clearly, at the critical point, $F$ significantly deviates from $1$, approaching the value $F=1+\langle\hat n\rangle$ at the critical point in the thermodynamic limit (shown by dashed-dotted lines).

\subsubsection{$P$-Representation}

The role of fluctuations in determining the emergent slow timescale can be proved 
by analyzing the Fokker-Planck equation in Eq.~(\ref{Eq:P-representation-evolution}).
As we previously discussed and demonstrated, all the Liouvillian eigenmatrices belonging to the $u_0$ sector (i.e., those which are invariant under any rotation, c.f. Appendix~\ref{Sec:Eigenmatrices_Wigner}) are unaffected by the parameter $\beta$. Thus, to investigate the dynamics of any initial state which is only a combination of matrices belonging to the $u_0$ sector, we can set $\beta=0$. The Fokker-Planck equation can be separated into radial and spatial components, i.e., $P(\alpha, t)=P(\theta, t)P(r,t)$, and $P(\theta, t)=P_{\rm ss}(\theta) = 1/2\pi$.
As such, the $P$-function for the field intensity $r$ acquires a similar form to that in the Eq.~(\ref{Eq:Psseq}) and
\begin{widetext}
\begin{equation}\label{Eq:Pr_no_eta}
\frac{\partial P(r,t)}{\partial t}=-\frac{1}{2}\frac{1}{r}\frac{\partial}{\partial
r}\left[r^2\left(A-\Gamma-B r^2\right)P(r,t)\right]+\frac{A}{4}\left(\frac{\partial^2}{\partial
r^2}+\frac{1}{r}\frac{\partial}{\partial
r}\right)P(r,t).
\end{equation}
\end{widetext}

In this case, the Fokker-Planck equation is an Ornstein-Uhlenbeck process, where the drift and diffusion terms correspond to the first and second terms in the right-hand side of Eq.~(\ref{Eq:Pr_no_eta}), respectively.  Moreover, the drift is not constant, but depends on the field intensity, i.e., the drift sign flips depending on whether the current intensity is larger or smaller than its mean value $\langle\hat n\rangle_{\rm ss}$ in the process, and which allows the system to evolve to its equilibrium.

\begin{figure}
    \centering
    \includegraphics[width=0.49 \textwidth]{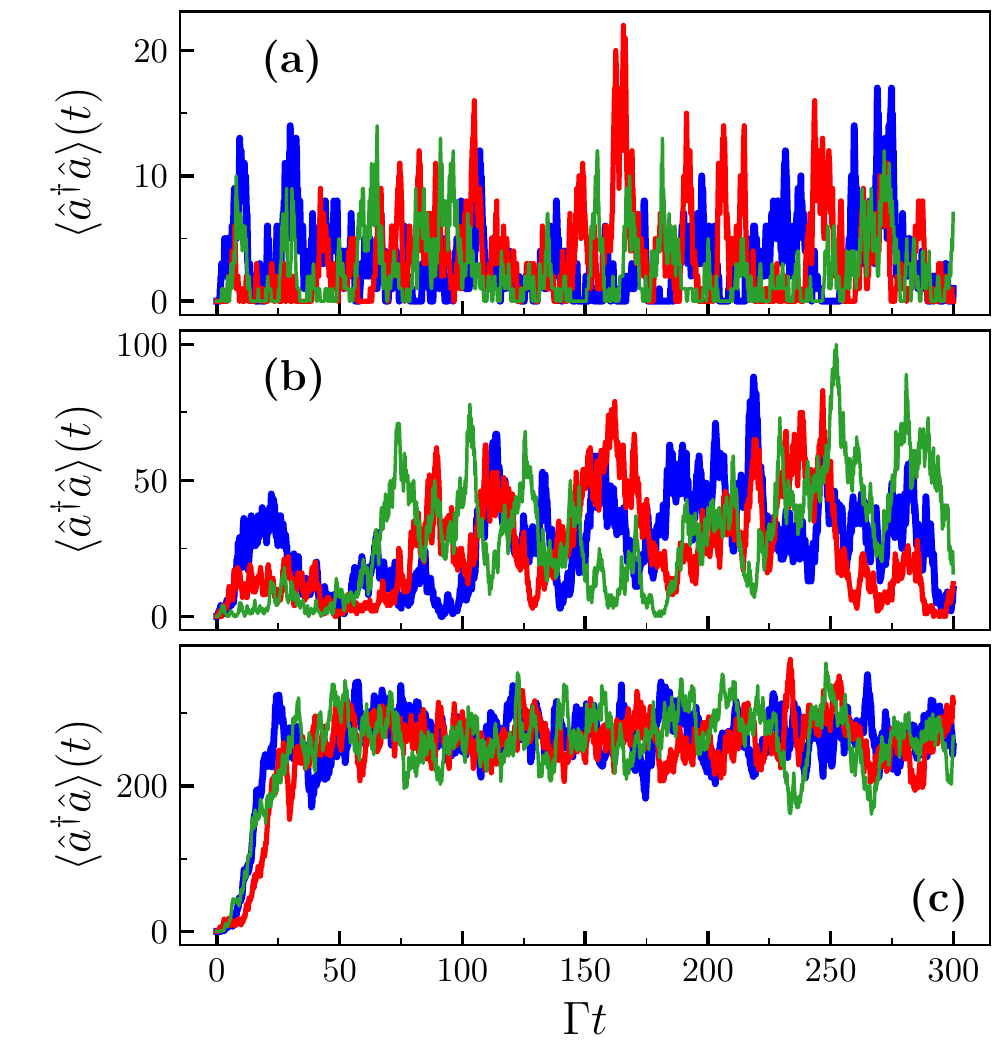}
    \caption{Photon number $\expec{\hat{a}^\dagger \hat{a}}$ along a single quantum trajectories for a state initialized in the vacuum for different gains (a) $A=3 \Gamma/4$, (b) $A=\Gamma$ (critical point), and (c) $A= 5 \Gamma/4$. Each line represents a different quantum trajectory.  Parameters: the saturation rate $B/\Gamma=10^{-1}$, $N=100$, and the frequency $\omega/\Gamma=1$ (these results are independent on $\omega$).}
    \label{fig:phot_chaos}
\end{figure}

We can identify three regimes: (1) well below the critical point ($A<\Gamma$), the mean photon number is small, thus the field intensity (photon number) just oscillates near its mean $\langle\hat n\rangle_{\rm ss}\approx0$ [see also Fig.~\ref{fig:phot_chaos}(a) for its visualization via quantum trajectories]. Also, the diffusion coefficient $A$ is small, resulting in a decreasing deviation from the mean with a decreasing gain.

(2) At the critical point ($A=\Gamma$), the mean photon number diverges in the thermodynamic limit $\langle\hat n\rangle_{\rm ss}\to \infty$. However, the rescaled photon number vanishes because $\langle\hat n\rangle_{\rm ss}/N\to 0$. We conclude that also the drift coefficient vanishes because $r(A-\Gamma-Br^2/N)\to0$ for sufficiently small $r$ in the thermodynamic limit. Consequently, possible fluctuations of the photon number around their mean induced by the drift are suppressed.
The typical timescale of the system is, thus, determined solely by the diffusion coefficient (proportional to the gain $A$).
Notably, the diffusion remains constant irrespective of the thermodynamic parameter $N$. As such, the typical timescale, which characterizes a system near the origin, can be seen as the time required from the system to explore the entire available configuration space, which becomes divergently large in the thermodynamic limit. 
For the finite-size systems, this diffusion-induced slowing down can be observed by trajectories initialized near the origin [as confirmed by Fig.~\ref{fig:phot_chaos}(b)].

(3) Finally, above the threshold ($A>\Gamma$), the drift becomes large again. Hence, the drift quickly drags the photon number to the proximity of its mean value, around which it remains to oscillate in the long-time limit [see Fig.~\ref{fig:phot_chaos}(c)]. The size of these fluctuations, however, becomes small with respect to the mean photon number.

\subsubsection{Quantum trajectories}

To visualize the effects of fluctuations in the full quantum simulation and a ``chaotic-like'' nature of the system at the critical point, we resort to quantum trajectories.
Quantum trajectories represent $\hat{\rho}(t)$ as the average over
several stochastic evolutions of a pure state wave
function. An in-depth discussion of the quantum trajectory formalism goes beyond the purpose of this article, and we refer the interested reader to, e.g., Refs.~\cite{Wiseman_BOOK_Quantum,DaleyAdvancesinPhysics2014}.
Suppose that we can monitor each time a quantum jump $\hat{L}_1$, $\hat{L}_2$, or $\hat{L}_3$ takes place.
Between two quantum jumps, the infinitesimal
evolution of the wave function  is dictated by a
non-Hermitian Hamiltonian \cite{MolmerJOSAB93,DalibardPRL92}, and
\begin{equation}\label{Eq:SSE}
\begin{split}
   \ket{\psi(t+dt)}&=\ket{\psi(t)} -i dt \hat{H}_{\rm eff} \ket{\psi(t)}, \\
   \hat{H}_{\rm eff}&= \omega \hat{a}^\dagger \hat{a} -\frac{i}{2} \left(\hat{L}_1^\dagger \hat{L}_1+\hat{L}_2^\dagger \hat{L}_2+\hat{L}_3^\dagger \hat{L}_3 \right).
\end{split}   
\end{equation}
This nonunitary evolution is interrupted by a random quantum jump of one of the operators $\hat{L}_j$, each one occurring at each interval ($t$, $t+dt$) with probability 
\begin{equation}
p_j   = dt \expec{\psi(t)|\hat{L}_j^\dagger \hat{L}_j|\psi(t)}.
\end{equation}
The occurrence of a quantum jump can be seen as a ``click'' of  a perfect detector \cite{Haroche_BOOK_Quantum}. In this case, the wave function instantaneously changes as
\begin{equation}
\ket{\psi(t+dt)} \propto  \hat{L}_j \ket{\psi(t)} ,
\end{equation}
where $dt$ is an infinitesimal time.

In Fig.~\ref{fig:phot_chaos} we plot the photon number along single quantum trajectories for different values of the gain $A$.
Before the critical point, $A<\Gamma$ in Fig.~\ref{fig:phot_chaos} (a), the signal is noisy and randomly passes between the vacuum and states with several photons.
Despite this fact, we see that the quantum trajectories change rapidly reaches its steady state value, around which it oscillates.
At the critical point, $A=\Gamma$ in Fig.~\ref{fig:phot_chaos} (b), we still see a very noisy signal, but in all the trajectories a long time is required before the system reaches states with a high number of photons. 
Finally, in Fig.~\ref{fig:phot_chaos} (c) where $A>\Gamma$, we observe a very different behavior: the system rapidly reaches a large photon number, and then oscillates around this value.

The intuition which we gain from this picture is that the critical point, similarly to the region $A<\Gamma$, is characterized by large oscillations. However, on a short timescale, the size of this oscillation is quite small [c.f. Fig.~\ref{fig:phot_chaos} (b)].
This implicates that the critical slowing down associated with the spectral collapse is due to both the chaotic nature of light (requiring to have large oscillations) and the fact that the size of the oscillations is comparatively small. 
Thus, it requires a large amount of time for any quantum trajectory to explore the entire space of accessible configurations. This picture confirms our predictions based on the $P$-representation analysis.

\section{Discussion and conclusions}
\label{Sec:Conclusions}

In this article, we analyzed the SLLM transition in the WGS regime, within the open quantum system framework provided by the spectral properties of the corresponding Liouvillian superoperator. 
Our analysis takes into account the nonequilibrium character of lasing and its quantum fluctuations, making this analysis more rigorous than previous studies, where the Landau theory and semiclassical analysis have been used.
Within the Liouvillian formalism, we demonstrated that the lasing transition in the SLLM is associated with a {\it Liouvillian spectral collapse}.

The spectral collapse indicates that, in one or more symmetry sectors [i.e., the Liouvillian eigenspaces whose dynamics is not interdependent due to the presence of the $U(1)$ symmetry], infinitely many eigenvalues retain their imaginary part, but collapse to zero in their real part.
As such, in the presence of symmetry breaking, the spectral collapse can be seen as the presence of infinitely many diabolical points [one for each symmetry sector of the $U(1)$ symmetry] of an infinite degeneracy.
Surprisingly, this spectral collapse does not signal the presence of an unphysical regime of the parameter space at larger gain rate $A$. Instead, it indicates a complete change in all the characteristics of the system.

This criticality, even in the absence of SSB, can be witnessed by the presence of a dynamical hysteresis. Remarkably, a semiclassical analysis, which neglects field fluctuations, completely misses this phenomenon. Contrary to ``usual'' multistability, which is suppressed by the presence of quantum fluctuations, this multistability is enabled by fluctuations. For example, in the seminal treatment of the Kerr resonator of Ref.~\cite{Drummond_JPA_80_bistability}, the semiclassical bistable solutions become a single density matrix, once quantum fluctuations are taken into account. This can be explained by the presence of quantum tunneling between the different semiclassical solutions~\cite{RiskenPRA87, Vogel_PRA_89_quasiprobability}, and a bistability emerges in the thermodynamic limit around the critical point of a first-order phase transition of the Kerr resonator~\cite{MingantPRA18_Spectral,LandaPRL20,LandaPRB20,LeBoitePRL13, BiondiPRA17,CasteelsPRA17-2, Foss-FeigPRA17,VicentiniPRA18, HuybrechtsPRA20}.

Our findings also prompt further investigations to analyze the Liouvillian spectral collapse within the framework of
dynamical phase transitions and even quantum chaos~\cite{Heyl2013,Fiderer2018,Cheraghi2020,Gutzwiller2007,Sedlmayr2019}.
Indeed, the presence of highly degenerate diabolical points describing nondecaying processes at different frequencies open perspectives for the study of the Loschmidt echoes and other indicators of dynamical
criticalities~\cite{Goussev2012,GORIN2006}.

The Landau theory and semiclassical analysis have also been used to study first-order lasing phenomena. For example, early studies of
lasers, with a low-saturation-intensity saturable absorber~\cite{Kazantsev1968,Salomaa1973}, sparked a number of
theoretical~\cite{Scott1975,Lugiato1978,Dembinski1978,Roy1979} and experimental~\cite{Okuda1977,Mortazavi2002} works in
characterizing first-order lasing transitions. Similarly,  first-order transitions in dye lasers were studied~\cite{Baczynsky1976,Schaefer1976,Marowsky1978}, and the analogy between the first-order phase transition and laser
threshold for multimode lasers led to the prediction of multicritical points~\cite{Hioe1981, Lett1981, Agarwal1982a,Agarwal1982b}. We plan to use the Liouvillian theory to investigate these classes of models.

\begin{acknowledgments}
The authors acknowledge the discussions with Alberto Biella, Simone Felicetti,
Alejandro Giacomotti, Simon Lieu, Carlos S\'anchez Mu\~noz, and Nathan
Shammah. 
The authors are grateful to the RIKEN Advanced Center
for Computing and Communication (ACCC) for the allocation of computational resources of the RIKEN supercomputer system (HOKUSAI BigWaterfall).
I.A. thanks the Grant Agency of the Czech Republic
(Project No.~18-08874S), and Project No.
CZ.02.1.01\/0.0\/0.0\/16\_019\/0000754 of the Ministry of
Education, Youth and Sports of the Czech Republic. A.M. is
supported by the Polish National Science Centre (NCN) under the
Maestro Grant No. DEC-2019/34/A/ST2/00081. F.N. is supported in part by: NTT Research,
Army Research Office (ARO) (Grant No. W911NF-18-1-0358),
Japan Science and Technology Agency (JST)
(via the CREST Grant No. JPMJCR1676),
Japan Society for the Promotion of Science (JSPS) (via the KAKENHI Grant No. JP20H00134
and the JSPS-RFBR Grant No. JPJSBP120194828),
the Asian Office of Aerospace Research and Development (AOARD) (via Grant No. FA2386-20-1-4069),
and the Foundational Questions Institute Fund (FQXi) via Grant No. FQXi-IAF19-06 Foundation.
\end{acknowledgments}

\appendix

\section{Liouvillian theory of Scully-Lamb laser in the weak gain saturation regime}

\begin{figure}
    \centering
    \includegraphics[width=0.7\linewidth]{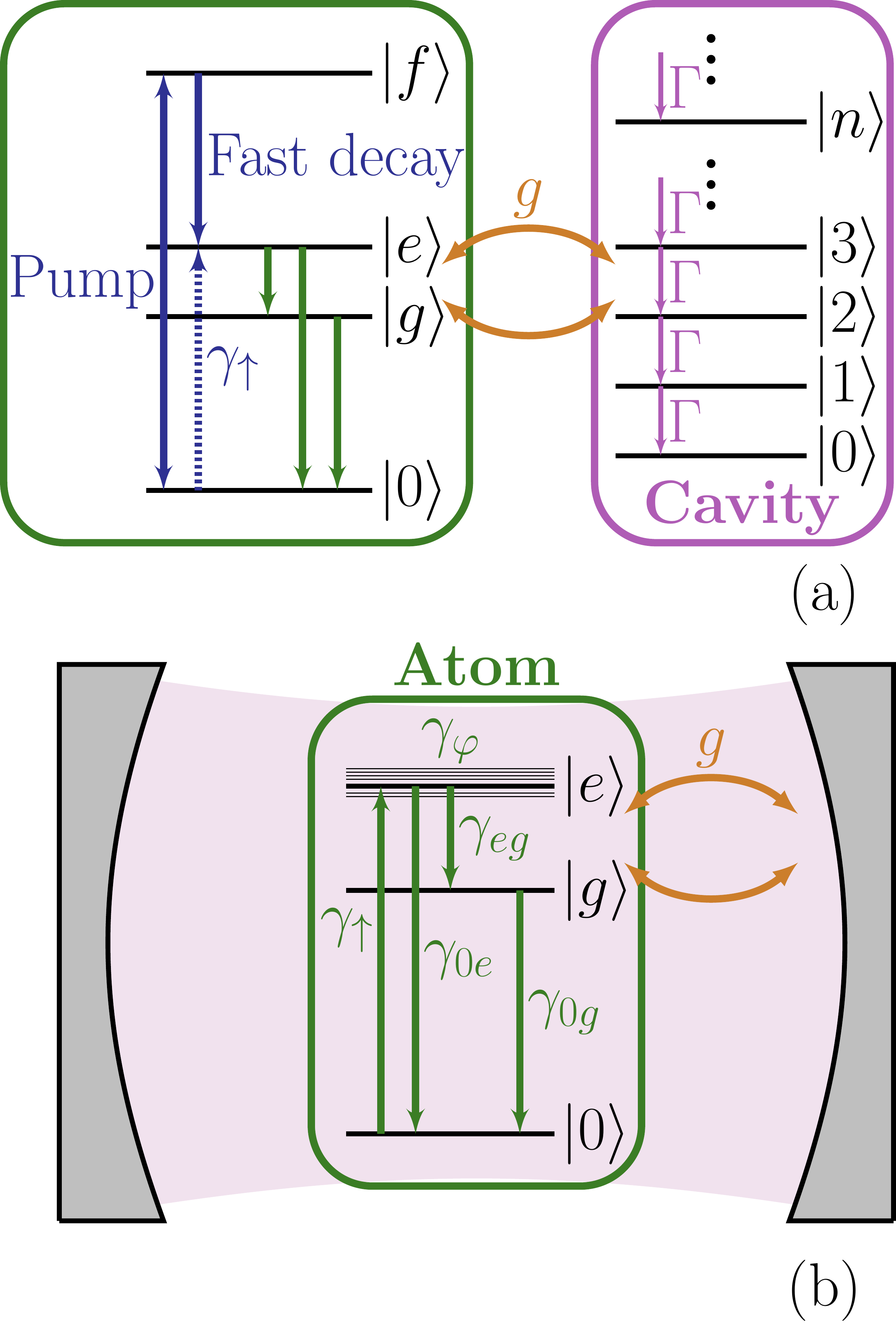}
    \caption{Pictorial representation of the Scully-Lamb laser model obtained via a driven four-level atoms system. (a) A beam of four-level atoms interacts with an optical cavity. An external pump drives the atoms from the lowest energy level $\ket{0}$ to the highest one $\ket{f}$. 
    (b) In the limit in which the decay from $\ket{f}$ is rapid, the emerging physics is that of an incoherently driven three-level atom interacting with a dissipative cavity, as detailed in the non-Lindbladian master equation in Eq.~(\ref{ME1}). The rates are explained in Table~\ref{table1}, and $g$ is the Jaynes-Cummings coupling strength between the atom and cavity. The broadening of $\ket{e}$
    schematically visualizes the effect of dephasing with a rate $\gamma_\varphi$.}
    \label{fig:scheme}
\end{figure}

\begin{table*}
\renewcommand{\arraystretch}{1.5}
\begin{tabular}{>{\centering\bfseries}m{1.5in} |> {\centering}m{1.9in} |>{\centering}m{1.4in} |>{\centering\arraybackslash}m{1.5in}}

    \hline\hline
    Master equations & \textbf{Process} & \textbf{(Super)operator} & \textbf{Rate} \\
    \hline
  \multirow{5}{1.5 in}{\centering atom-cavity-field LME in Eq.~(\ref{ME1})}
  & atomic decays (emissions)
  & $\DD[\hat{\sigma}_{p,q} =\ket{p}\bra{q}]$
  & $\gamma_{p,q}$ \\

  & atom pumping
  & $\DD[\hat{\sigma}_{e,0}]$
  & $\gamma_\uparrow$ \\
  & atom dephasing
  & $\DD[\hat{\sigma}_{e,e}- \hat{\sigma}_{g,g}]$
  & $\gamma_\varphi$ \\
  & cavity decay
  & $\DD[\hat{a}]$
  & $\Gamma$ \\
  & atom-cavity interaction
  & $\hat{H}_{\rm JC}$ in \eqref{Eq:JC}
  & $g$ \\
\hline
  \multirow{4}{1.5 in}{\centering cavity-field non-LME in Eq.~(\ref{Eq:SL_ME})}
  & linear gain
  & $\DD[\hat{a}^\dagger]$
  & $A$ [Eqs. (\ref{Eq:fast_dephasing}) or (\ref{Eq:broadened_lifetime})]\\
  
  & dephasing
  & $\DD[\hat{a}\hat{a}^\dagger]$
  & $B_1$ \\
  

  & gain saturation
  & ${\cal K}[\hat{a}]$
  &
  $B_2$ \\

  & cavity decay
  & $\DD[\hat{a}]$
  & $\Gamma$ \\

\hline
  \multirow{3}{1.5 in}{\centering cavity-field LME in Eq.~(\ref{Eq:SL_Liouv})}
  & nonlinear gain
  & $\DD \big[\hat{a}^\dagger\big(1-\frac{B}{2A}\hat{a}\hat{a}^\dagger\big)\big]$
  & $A$ \\
  
  & dephasing
  & $\DD [\hat{a}^\dagger \hat{a}]$
  & $3B/4$ \\  
  

  & cavity decay
  & $\DD [\hat{a}]$
  & $\Gamma$ \\

\end{tabular}
\caption{Master equations for the Scully-Lamb model in their
Lindblad (LME) and non-Lindblad (non-LME) forms (recall that $\hat{\sigma}_{p,q}=\ket{p}\bra{q}$).} \label{table1}
\end{table*}

\label{Sec:Liouvillian_theory_full}

\subsection{Non-Lindbladian master equations for the Scully-Lamb laser model}
\label{Sec:Non_Lindblad_ME}
Here, we consider the minimal model to obtain the Scully-Lamb laser
master equation, inspired by the derivation of Yamamoto and Imamo\v{g}lu in Ref.~\cite{YamamotoBook}. 

We consider a beam of four-level atoms whose states, in order of ascending energy, are $\{\ket{0},\,  \ket{g}, \, \ket{e}, \, \ket{f}\}$. The operators $\hat{\sigma}_{p,q}\equiv \ket{p}\bra{q}$ indicate the atomic transitions from $\ket{q}$ to $\ket{p}$.
Spontaneous decays enable the system to decrease its energy via the transitions
$\ket{q} \to \ket{p}$, if $\ket{q}$ has a higher energy than $\ket{p}$, described by a set of Lindblad dissipators $\gamma_{p,q}\DD[\hat{\sigma}_{p,q}]$ [see Table~\ref{table1} and Fig.~\ref{fig:scheme}(a)].

A cavity of frequency $\omega_c$ is at resonance with the $\ket{g} \leftrightarrow \ket{e}$ transition, while all the other transitions are far-off resonance. 
Under this condition, light and matter can exchange a single excitation, and
the coherent part of the system is described by the Jaynes-Cummings Hamiltonian \cite{kockum2019}
\begin{equation}\label{Eq:JC}
    \hat{H}_{\rm JC} = \omega_c\hat{a}^\dagger \hat{a}+ \frac{\omega_q}{2} \hat{\sigma}_{z} + g \left(\hat{a} \hat{\sigma}_{e,g} + {\rm h.c.} \right),
\end{equation}
where $\hat{\sigma}_{z}=\hat{\sigma}_{e,e} - \hat{\sigma}_{g,g}$.
The cavity photons are lost via $\Gamma \DD[\hat{a}]$; the term $\gamma_\varphi \DD[\hat{\sigma}_{z}]$ describes dephasing due to, e.g., collisions between different species of atoms in the laser, which destroy the coherence between $\ket{e}$ and $\ket{g}$ induced by the cavity field.

An external coherent field continuously pumps (excites) the atoms from their ground level $\ket{0}$ to the uppermost level $\ket{f}$ and vice versa. 
Let us assume that the drive is strong and that the decay rate $\gamma_{e,f}$, describing the transition $\ket{f} \to \ket{e}$, is faster than all the other rates of the system. Then, one can trace out the $\ket{f}$ degree of freedom, obtaining a non-Hermitian effective pumping $\gamma_{\uparrow}\DD[\ket{e}\bra{g}]$ [c.f. Fig.~\ref{fig:scheme}(a)].

Altogether, the evolution of the resulting three-level system coupled to the optical cavity is governed by this non-Lindbladian atom--cavity-field cavity master equation ($\hbar=1$):
\begin{eqnarray}\label{ME1}
   &&\frac{d}{dt}\rhot=-i[H_{\rm JC},\rhot]+ \left(\Gamma \DD[\hat{a}] + \gamma_{ge}\DD[\hat{\sigma}_{ge}] + \gamma_{0e} \DD[\hat{\sigma}_{0e}] \right. \nonumber\\
                    & &\quad \left. + \gamma_{0g} \DD[\hat{\sigma}_{0g}] + \gamma_{\uparrow} \DD[\hat{\sigma}_{e0}] + \gamma_\varphi \DD [\hat{\sigma}_{ee} - \hat{\sigma}_{gg}] \right) \rhot,
\end{eqnarray}
as sketched in Fig.~\ref{fig:scheme}(b).

An effective laser master equation for the cavity alone can be obtained in the two limits:
\begin{enumerate}
    \item The dephasing process dominates the dynamics [i.e., the fast dephasing case $\gamma_\varphi \gg \gamma_{0g}\gg (\gamma_{0e} + \gamma_{ge})$];
    \item An idealized system where $\gamma_{0e} =\gamma_{0g} =\gamma$, while $\gamma_{ge} = \gamma_\varphi=0$ [i.e., a lifetime broadened laser].
\end{enumerate}
In both cases, by tracing out the atomic degrees of freedom, a non-Lindblad-form for the reduced density matrix $\rhot\equiv \rhot_{\rm cav}$ of the cavity field can be obtained:
\begin{equation}\label{Eq:SL_ME}
\begin{split}
   \frac{d}{dt}\rhot &=-i [\omega_c \hat{a}^\dagger \hat{a}, \rhot] + \Big\{\Gamma \DD[\hat{a}]
   +A \DD [a^{\dagger}] \\
   &  +B_1\DD[\hat{a} \hat{a}^{\dagger}]-B_2{\cal K}[\hat{a}] \Big\} \rhot,
\end{split}   
\end{equation}
where $A$, $B_1$, and $B_2$ are the gain, dephasing, and gain saturation rates. 
In the fast dephasing case, they are:
\begin{equation}\label{Eq:fast_dephasing}
A=\frac{4\gamma_\uparrow g^2}{\gamma_{\rm tot} (\gamma_{0e}+\gamma_{ge})}, \quad
B_1=\frac{A^2}{4\gamma_\uparrow}, \quad
B_2=  B_1,
\end{equation}
where $\gamma_{\rm tot}=\gamma_\varphi+\gamma_{0e}+\gamma_{ge}+\gamma_{0g}$.
In the lifetime broadened laser, instead, one has
\begin{equation}\label{Eq:broadened_lifetime}
    A=\frac{\gamma_\uparrow g^2}{ \gamma^2}, \quad
B_1=\frac{3 A^2}{2 \gamma_\uparrow}, \quad
B_2=\frac{2B_1}{3}.
\end{equation}
The superoperator $\mathcal{K}$ is not a Lindblad dissipator, and it makes \eqref{Eq:SL_ME} non-Lindbladian, with
\begin{equation}\label{K}
  {\cal K}[\hat{a}] \rhot = \hat{a}^\dagger \{\hat{a} \hat{a}^\dagger, \rhot \}\hat{a} -  \{(\hat{a} \hat{a}^\dagger)^2, \rhot \},
\end{equation}
where $\{\hat{A}, \hat{B}\}=\hat{A}\hat{B}+\hat{B}\hat{A}$.

In the following, and for the sake of simplicity, we focus on the lifetime broadened laser (the case 2) by choosing $B=B_1=B_2$. Similar results can be obtained for case (1) upon appropriate rescaling of the parameters.

\subsection{Lindblad form in the weak-gain saturation limit}
\label{APP:WGS}
The master equation in \eqref{Eq:SL_ME} can be obtained in several ways~\cite{ScullyLambBook, OrszagBook}.
In particular, \eqref{Eq:SL_ME} is an approximation of the original Scully-Lamb equation in Ref.~\cite{Scully1967} valid up to \textit{third-order} corrections in the field amplitude.

Although trace preserving, \eqref{Eq:SL_ME} does not conserve the complete positivity of the density matrix (i.e., it can lead to a negative probability distribution of the eigenstates of $\rhot$~\cite{HenkelJPB07}). 
Since all Lindblad master equations are completely positive trace-preserving (CPTP) maps, \eqref{Eq:SL_ME} cannot be brought to a Lindblad form. 
However, by ignoring terms
of \textit{second order} in $\hat{a}\hat{a}^\dagger B/(2A)$ [which corresponds to the WGS limit in \eqref{Eq:Weak-gain-saturation}] Eq.~(\ref{Eq:SL_ME}) becomes \cite{Gea1998,Arkhipov2019}
\begin{equation}\label{Eq:SL_Liouv}
\begin{split}
    \frac{\de}{\de t} \rhot &
    -i\left[\omega_c \hat{a}^\dagger \hat{a}, \rhot \right]  + \sum_{j=1}^{3} \DD[\hat{L}_j] \rhot,
    \end{split}
\end{equation}
where $\hat{L}_j$ are exactly the jump operators obtained in \eqref{Eq:L123} (see also Table~\ref{table1}).
As demonstrated in Ref.~\cite{HenkelJPB07} for a micromaser model, such a Lindblad form can also by obtained by an expansion in the light-matter coupling strength that explicitly preserves the Lindblad form of the master equation.

Equations (\ref{Eq:SL_ME}) and (\ref{Eq:SL_Liouv}) must be taken with a grain of salt. Since they are valid in the WGS limit, they cannot be used to describe processes involving a high number of excitations.
Consider, e.g., an initial Fock state $\hat{\rho}(t=0)=\ket{m}\bra{m}$. If $m$ is small enough and $B\ll A$, the leading term in $\hat{L}_1$ is $\sqrt{A}\hat{a}^\dagger$, and the dynamics rapidly converges towards its steady state, where $A$ and $B$ compete. If, however, $m^2 \gg A/B$, the dynamics is dominated by the $B$ term which is unbounded, and the system cannot converge anymore to the ``true'' steady state, and the number of photons diverges towards infinity. In other words, there exist values of $m$ for which the states become nonphysical.
From a Liouvillian point of view, the SLLM is unbounded
and, increasing the cutoff, spurious eigenvalues from the
high-excited eigenmatrices of the Liouvillian (which do not
respect the WGS conditions) produce quasi-zeros of the spectrum.

In both simulating the dynamics and studying the Liouvillian
spectrum, this problem plays a very minor role, since spurious eigenvalues can be easily eliminated. Indeed,
when discussing the dynamics of the system, we verified that all
the solutions are dynamically stable (i.e., the presence of a
diverging dynamics does not affect the steady state). Moreover, in
the same way in which the low-lying part of the dynamics is
unaffected by the presence of the diverging dynamics for a
high-photon number, the Liouvillian eigenstates converge to a
well-defined value, even when spurious eigenvalues are present.
Thus the unphysical eigenvalues and eigenmatrices can be easily
discarded \textit{a posteriori}.

A more detailed analysis of the problem and a deeper discussion of the validity of this approximation can also be found in
Ref.~\cite{WangPRA73}.

\begin{figure*}
    \centering
    \includegraphics[width=0.95 \textwidth]{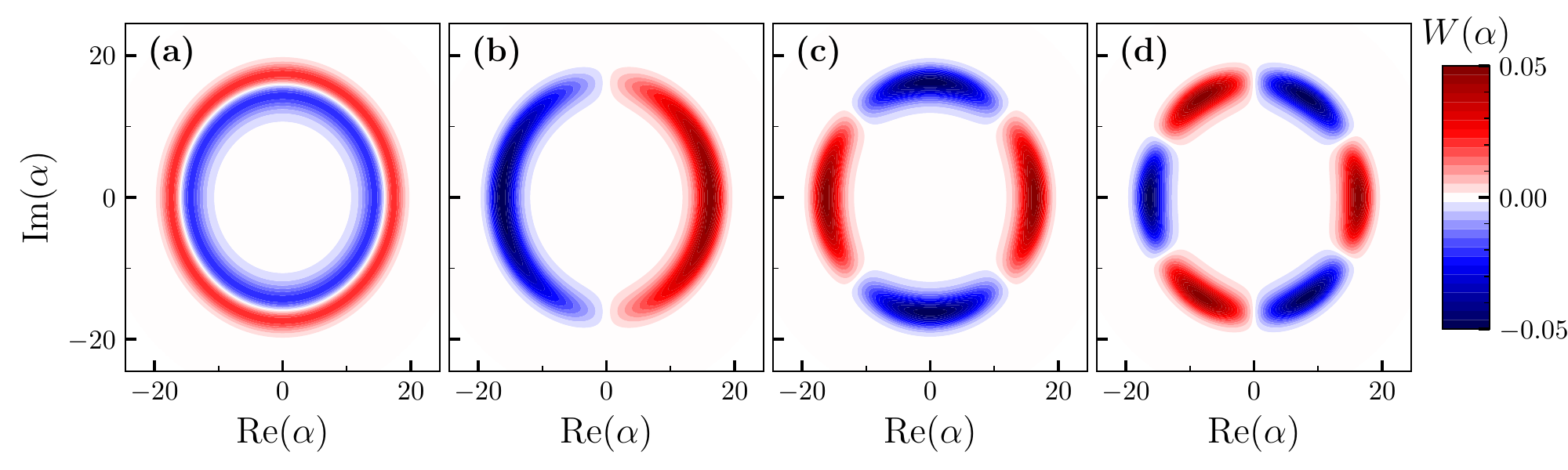}
    \caption{The Wigner representation in \eqref{Eq:Winger_repr} capturing the physical meaning of the eigenmatrices $\eig{j}^{(k)}$ of each symmetry sector. (a) $\eig{1}^{(0)}$ is invariant under any rotation and, thus, describes purely exponentially decaying processes; (b) $\eig{0}^{(1)}$ has a $\pi$ symmetry and describes processes of frequency $\omega$, while (c) $\eig{0}^{(2)}$ and (d) $\eig{0}^{(3)}$ describe processes at frequencies $2 \omega$ and $3\omega$ being characterized by $\pi/2$ and $\pi/3$ symmetries, respectively. The negativity of the Wigner representation originates from the fact that all the eigenmatrices, except the steady state $\sss$, are traceless. Parameters: $A/\Gamma=1.25$, $B/\Gamma=10^{-1}/N$, and $N=50$.}
    \label{fig:wigner_eigenstates}
\end{figure*}

\subsection{Physical meaning of the $U(1)$ symmetry and eigenmatrices
of the Liouvillian}
\label{Sec:Eigenmatrices_Wigner}
Given \eqref{Eq:U_1_superoperator}, we can immediately determine
various properties of the eigenmatrices belonging to different
symmetry sectors. In the number (Fock) basis, the eigenmatrix
$\eig{i}$ reads
\begin{equation}\label{rho_i}
    \hat\rho_i=\sum_{m,n} c_{m,n}\ket{m}\bra{n}.
\end{equation}
By combining Eqs. (\ref{Eq:U_1_superoperator})~and~(\ref{rho_i}),
one obtains
\begin{equation}\label{u_i}
\begin{split}
        \mathcal{U}\eig{i} &= \sum_{m,n} c_{m,n} \exp\left( - i \phi \hat{a}^\dagger \hat{a}\right)  \ket{m}\bra{n} \exp\left( i \phi \hat{a}^\dagger \hat{a}\right)  \\
        & =\sum_{m,n} c_{m,n} \exp\left[i \phi(n-m)\right]\ket{m}\bra{n}= u_i \eig{i}.
\end{split}
\end{equation}
We conclude that $\exp\left[i \phi(n-m)\right]$ must be a constant and,
therefore, it holds
\begin{equation}\label{Eq:condition_symmetry_U(1)}
    \eig{j}^{(k)}= \sum_{m} c_{m} \ket{m}\bra{m-k} \, ,
\end{equation}
for any constant integer $k$. In other words,
$\eig{i}$ must be an operator containing elements only on one
diagonal, and different symmetry sectors should occupy different
upper and lower diagonals.
From a practical point of view, by introducing a cutoff $N$ the
$U(1)$ symmetry becomes $Z_N$. 
Within this representation, $k \in
[0, N-1]$.

We recall that the eigenmatrices have, by themselves, no physical
meaning but they need to be combined with $\sss$ in order to
represent a physical process \cite{MingantPRA18_Spectral}.
While $\sss+\sum_j c_j \eig{j}^{(0)}$ is well-defined state, for a generic $k$ one needs to consider
\begin{equation}
\rhot=\sss+\sum_j c^{+}_j \left(\eig{j}^{(k)} + \eig{j}^{(-k)} \right) + c^{-}_j \left(\eig{j}^{(k)} - \eig{j}^{(-k)} \right), 
\end{equation}
because the $\eig{j}^{(k)}$ are non Hermitian.

Nevertheless, these eigenmatrices and their symmetries have
a clear physical interpretation. 
They can be visualized by their Wigner representation, i.e.
\begin{equation}\label{Eq:Winger_repr}
W_j^{(k)}(\alpha) =\frac{1}{\pi}
\operatorname{Tr}\left[\hat{D}_{\alpha} \exp\left(i \pi \hat{a}^{\dagger}
\hat{a}\right) \hat{D}_{\alpha}^{\dagger} \left(\eig{j}^{(k)}+\eig{j}^{(-k)} \right) \right],
\end{equation}
where $\hat{D}_{\alpha}=\exp{(\alpha \hat{a}^\dagger - \alpha^{*}
\hat{a})}$ is the displacement operator. Notice that, for a
generic density matrix, the Wigner representation is the standard
Wigner function describing a quasiprobability. For the
eigenmatrices under consideration, this is not the case since,
except the steady-state, they are traceless. Moreover, the
negativities of these Wigner representations of the eigenmatrices
are not indicators of the ``quantumness'' of the state.

In Fig.~\ref{fig:wigner_eigenstates}, we plot different
eigenmatrices of the Liouvillian, corresponding to different
symmetry sectors. Each sector is characterized by a different
structure. A matrix corresponding to $u_0$ must be invariant under
any rotation and, therefore, it has a circular symmetry in its
Wigner representation [c.f. Fig.~\ref{fig:wigner_eigenstates}(a)].
As it can be seen in Fig.~\ref{fig:wigner_eigenstates}(b), a state
belonging to $u_1$ must be invariant upon a $\pi$ rotation (up to
the minus sign). Similarly, for
Figs.~\ref{fig:wigner_eigenstates}(c) and \ref{fig:wigner_eigenstates}(d), the rotation becomes
$\pi/2$ and $\pi/3$, respectively.

We conclude that $u_0$ describes all those phase-independent exponentially decaying processes.
The sector $u_1$ captures all those processes that have a single lobe in the phase
space. This means that they capture the dynamics towards the
steady state of those processes rotating at a frequency $\omega$.
Similarly, $u_k$ is characterized by $k$
identical lobes in the phase space, and  it rotates at the
frequency $k \omega$.

\section{The effect of phase fluctuations}
\label{Sec:Quantum_Fluctuations}

To understand why the semiclassical approximation (which often correctly captures criticality in other single-cavity models \cite{BartoloPRA16,CasteelsPRA17-2,SavonaPRA17}) fails for $\eta \neq 0$, we consider here the fluctuations induced by dephasing.
The semiclassical approximation assumes the trivial relation $n=|\alpha|^2$. However, for
symmetry reason, in the full model we have that
$\expec{\hat{a}}_{\rm ss}=\exp(i \phi) \expec{\hat{a}}_{\rm ss} =
0$. As such, the value $\expec{\hat{a}}$ can be different from
zero only in the dynamics towards the steady state or in the thermodynamic limit, where the $U(1)$ SSB takes place. 
Equivalently, the semiclassical approximation is mixing dynamical
quantities with steady-state ones, and it is therefore valid only
if $\eta=0$ and coherence can be maintained for an ``infinite''
time.

Using the previously introduced counting-trajectory formalism, we can characterize
the effect of $\eta$ on a coherent state. If no quantum jump happens, the system 
$\hat{H}_{\rm eff}$ contains the term $-\eta/2 (\hat{a}^\dagger)^2
\hat{a}^2 $, which acts as an imaginary Kerr nonlinearity,
inducing dephasing and changing the shape of a coherent state.
Similarly,  a quantum jump $\hat{L}_2$ destroys the coherence of a state (i.e., the
coherent state is \textit{not} an eigenstate of $\hat{a}^\dagger \hat{a}$).
These are not the fluctuations of the photon number discussed above.

\begin{figure}
    \centering
    \includegraphics[width=0.49 \textwidth]{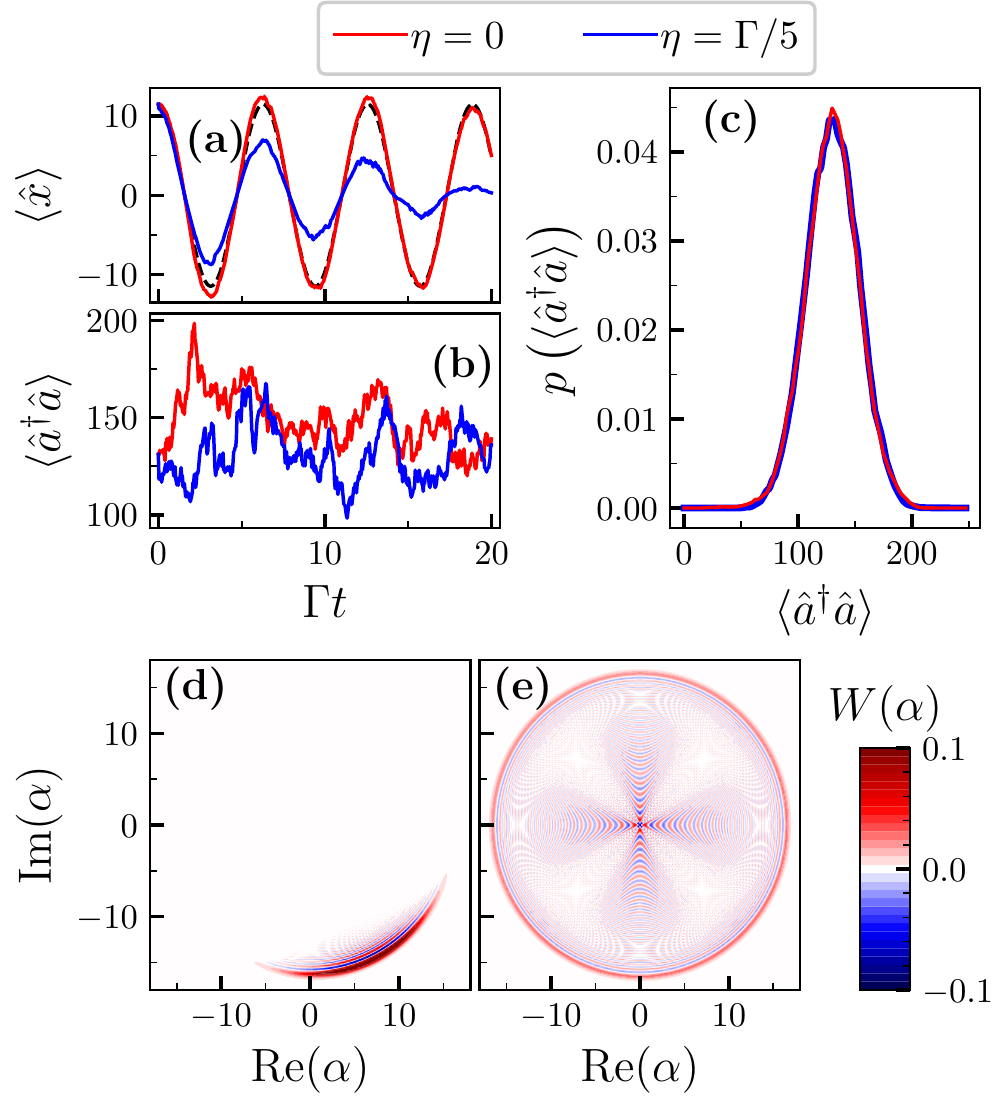}
    \caption{Single counting quantum trajectory for a state initialized in a coherent state $\ket{\alpha}$, where $|\alpha|^2 = \expec{\hat{a}^\dagger \hat{a}}_{\rm ss}$, for $ \eta=0$ (red curves) and $ \eta = \Gamma /5$ (blue curves). (a) Time evolution of the $\hat{x}$ quadrature. (b) Photon number $\expec{\hat{a}^\dagger \hat{a}}$ as a function of time. (c) Probability distribution of $\expec{\hat{a}^\dagger \hat{a}}$ for a single quantum trajectory. To collect a sufficient statistics, the trajectory was run for a total time $t=10^{4} \times \Gamma$. (d) Wigner function at time $t=20 \Gamma$ for $ \eta=0$. (e) Wigner function at time $t=20 \Gamma$ for $ \eta=\Gamma /5$. Parameters: gain $A/\Gamma$=5/4, saturation rate $B/\Gamma=10^{-1}$, $N=100$, and frequency $\omega/\Gamma=1$. }
    \label{Fig:Trajectory_counting}
\end{figure}

\begin{figure}
    \centering
    \includegraphics[width=0.49 \textwidth]{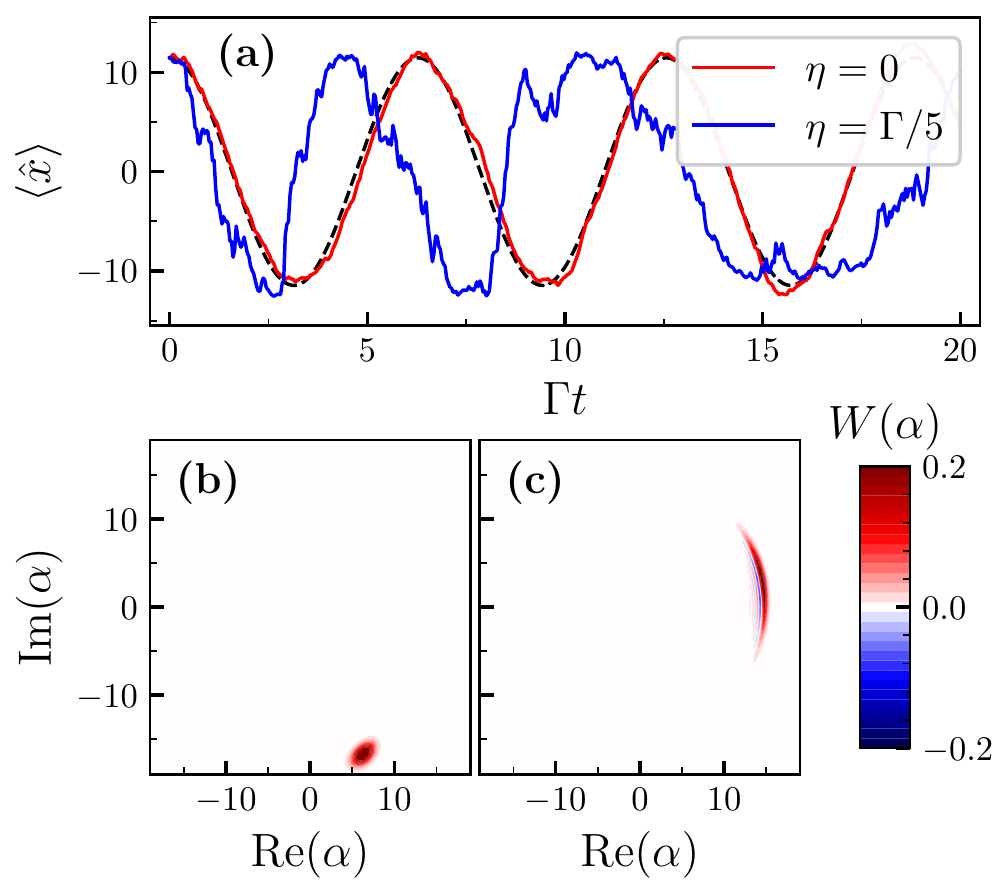}
    \caption{Single homodyne quantum trajectory for a state initialized in a coherent state $\ket{\alpha}$, where $|\alpha|^2 = \expec{\hat{a}^\dagger \hat{a}}_{\rm ss}$. (a) Time evolution of the $\hat{x}$ quadrature for $ \eta=0$ (red curve) and $ \eta = \Gamma /5$ (blue curve). (b) Wigner function at time $t=20 \Gamma$ for $ \eta=0$. (c) Wigner function at time $t=20 \Gamma$ for $ \eta=\Gamma /5$. Parameters: gain $A/\Gamma=5/4 $, gain saturation $B/\Gamma=10^{-1}/N$, $N=100$, and frequency $\omega/\Gamma=1$. }
    \label{fig:homodyne}
\end{figure}

We plot the results of such a quantum trajectory in
Fig.~\ref{Fig:Trajectory_counting} for an initial coherent state.
In Fig~\ref{Fig:Trajectory_counting}(a), we show the expectation value of
$\hat{x}=(\hat{a}+\hat{a}^\dagger)/2$ both for $
\eta=0$ (corresponding to the standard SLLM, red curve) and
$ \eta \neq 0$ (i.e., the generalized SLLM, blue
curve). In both cases, $\expec{\hat{x}}$ oscillates at a frequency $\omega$. While for $\eta =0$ the oscillations are long-lived, for $
\eta \neq 0$ they rapidly approach zero. This is
corroborated by the Wigner functions at time $t=20 \, \Gamma$
in Figs.~\ref{Fig:Trajectory_counting}(d)~and~\ref{Fig:Trajectory_counting}(e). While in the
standard Scully-Lamb the system retains a well-defined phase,
in the presence of dephasing $ \eta \neq 0$ the state
rapidly loses any coherent-like feature. Finally, in Fig.~\ref{Fig:Trajectory_counting}(c)
we plot $p(\expec{\hat{a}^\dagger
\hat{a}})$, i.e., the probability distribution that the photon number attain a certain value along a single very long quantum trajectory, both with or without additional dephasing. Also, this
distribution is independent of the value of $\eta$.

The results of Fig.~\ref{Fig:Trajectory_counting} seem to
indicate that, for a very long trajectory, the system would
completely lose its coherence, thus resulting in a state which
does not oscillate anymore. One would be tempted to interpret the $P$-function solution as a superposition of states which have completely lost their coherence. This picture, however, is due to the particular assumptions leading to \eqref{Eq:SSE}. Indeed, as Eqs.~(\ref{Diffusion_eq}) and (\ref{D}) imply, the dephasing $\eta$ determines a {\it{random-walk}} phase diffusion of the field.
To illustrate this, we can consider a different kind of unraveling (e.g., a diffusive homodyne-like quantum
trajectory \cite{Wiseman_BOOK_Quantum}), the result of the
statistical unraveling then would be completely different.
In this case, a reference field of intensity $\beta$ is mixed with the output field of each jump operator. Consequently, the jump operators become $\hat{L}_j (\beta)= \hat{L}_j + \beta_j$, while the Hamiltonian reads
\begin{equation}
    \hat{H}_{\rm eff}(\beta) = \hat{H}_{\rm eff} - i \sum_j    \beta_j \left(\hat{L}_j - \hat{L}_j^\dagger  \right).
\end{equation}
In the ideal ``homodyne'' trajectory limit $\beta_j\to \infty$, the detector continuously reads a signal. The effect of the measurement on the system, however, is minimal, since each quantum jump is largely due to the presence of the local oscillator $\beta_j$.

We plot
these results in Fig.~\ref{fig:homodyne}. This time, the effect of
decoherence can be interpreted as random displacement (walk) which a
coherent state undergo [Fig.~\ref{fig:homodyne}(a)]. As such, the
dynamics of a single quantum trajectory always retain a nonzero phase [c.f. Figs.~\ref{fig:homodyne}(b)~and~\ref{fig:homodyne}(c)].
In this case, the $P$-function would result from the superposition
of several coherent states with different phases and amplitudes.

We conclude that the photon number and its fluctuation depend only on the gain $A$, saturation $B$, and dissipation $\Gamma$. We confirm that $\eta$ plays no role in determining them, but induces decoherence, which can be interpreted in different ways: (i) in a counting trajectory approach, the decoherence changes the shape of an otherwise coherent-like state; (ii) in a homodyne trajectory, $\eta$ randomly displaces the phase of the wave function, preventing long-term coherence.
Neither dephased states nor random fluctuations can be described as a single coherent state. Thus, the semiclassical approach based on \eqref{Eq:Semiclassical_solution} assigns to $\eta$ the same role as $\Gamma$, \textit{wrongly} predicting the presence of SSB in the generalized SLLM.


%

\end{document}